\pdfoutput=1

\newif\ifarXiv
\arXivtrue 

\ifarXiv

\documentclass[
 aps, 
 pre,superscriptaddress,showpacs, 
 showkeys, 
 numerical,
 nofootinbib
]{revtex4-1} 

\else

\documentclass[NETN]{stjour}

\articletype{Methods}

\supportinginfo{www.github.com/pwollstadt/IDTxl}

\received{}
\accepted{}
\published{}
\handlingeditor{}

\setdoi{}

\fi

\newcommand{\nn}{\nonumber}
\newcommand{\ie}{{i.e.}}
\newcommand{\eg}{{e.g.}}
\newcommand{\fig}[1]{Figure~\ref{fig:#1}}
\newcommand{\tb}[1]{Table~\ref{tb:#1}}

\newcommand{\eq}[1]{(\ref{eq:#1})}

\usepackage{graphicx}
\graphicspath{{images/}}
\usepackage{booktabs,multirow} 
\usepackage{microtype}
\usepackage{amsmath,amssymb,bm,siunitx}
\usepackage{enumitem}
\usepackage{lineno}

\newcommand{\paperTitle}{Large-scale directed network inference with multivariate transfer entropy and hierarchical statistical testing}
\newcommand{\shortTitle}{Large-scale network inference with multivariate transfer entropy}
\newcommand{\theKeywords}{neuroimaging, directed connectivity, effective network, multivariate transfer entropy, information theory, nonlinear dynamics, statistical inference, nonparametric tests}

\ifx\pdfoutput\undefined
\else
\usepackage[bookmarks=true,colorlinks=false,citecolor=black,linkcolor=black,breaklinks=true,pdftex]{hyperref}
\hypersetup{
      	pdfauthor = {Leonardo Novelli},
        pdftitle = {\paperTitle},
  	    pdfsubject = {},
      	pdfkeywords = {\theKeywords},
        pdfcreator = {LaTeX with hyperref package},
  	    pdfproducer = {pdftex},
  	    pdfstartview = {FitH}}
\fi

\begin{document}


\ifarXiv


\title{\paperTitle}

\author{Leonardo Novelli}
\thanks{First authors contributed equally to this work.}
\affiliation{Centre for Complex Systems, Faculty of Engineering and IT, The University of Sydney, Sydney, Australia}
\author{Patricia Wollstadt}
\thanks{First authors contributed equally to this work.}
\affiliation{Honda Research Institute Europe, Offenbach am Main, Germany}
\author{Pedro Mediano}
\affiliation{Computational Neurodynamics Group, Department of Computing, Imperial College London, London, United Kingdom}
\author{Michael Wibral}
\affiliation{Campus Institute for Dynamics of Biological Networks, Georg-August University, G\"ottingen, Germany}
\author{Joseph T. Lizier}
\affiliation{Centre for Complex Systems, Faculty of Engineering and IT, The University of Sydney, Sydney, Australia}

\date{\today}

\else


\title[\shortTitle]{\paperTitle}

\author[]
{Leonardo Novelli\affil{1,*},
Patricia Wollstadt\affil{2,*},
Pedro Mediano\affil{3},
Michael Wibral\affil{4},\\
\and Joseph T. Lizier\affil{1}}

\affiliation{1}{Centre for Complex Systems, Faculty of Engineering and IT, The University of Sydney, Sydney, Australia}

\affiliation{2}{Honda Research Institute Europe, Offenbach am Main, Germany}

\affiliation{3}{Computational Neurodynamics Group, Department of Computing, Imperial College London, London, United Kingdom}

\affiliation{4}{Campus Institute for Dynamics of Biological Networks, Georg-August University, G\"ottingen, Germany}

\affiliation{*}{First authors contributed equally to this work.}

\correspondingauthor{Leonardo Novelli}{leonardo.novelli@sydney.edu.au}



\fi


\begin{abstract} 
Network inference algorithms are valuable tools for the study of large-scale neuroimaging datasets.
Multivariate transfer entropy is well suited for this task, being a model-free measure that captures nonlinear and lagged dependencies between time series to infer a minimal directed network model. Greedy algorithms have been proposed to efficiently deal with high-dimensional datasets while avoiding redundant inferences and capturing synergistic effects.
However, multiple statistical comparisons may inflate the false positive rate and are computationally demanding, which limited the size of previous validation studies.
The algorithm we present---as implemented in the IDTxl open-source software---addresses these challenges by employing hierarchical statistical tests to control the family-wise error rate and to allow for efficient parallelisation. The method was validated on synthetic datasets involving random networks of increasing size (up to \num{100} nodes), for both linear and nonlinear dynamics.
The performance increased with the length of the time series, reaching consistently high precision, recall, and specificity ($>98\%$ on average) for \num{10000} time samples. Varying the statistical significance threshold showed a more favourable precision-recall trade-off for longer time series.
Both the network size and the sample size are one order of magnitude larger than previously demonstrated, showing feasibility for typical EEG and MEG experiments.
\end{abstract}

\ifarXiv

\maketitle 

\fi

\section{Introduction}
The increasing availability of large-scale, fine-grained datasets provides an unprecedented opportunity for quantitative studies of complex systems. Nonetheless, a shift towards data-driven modelling of these systems requires efficient algorithms for analysing multivariate time series, which are obtained from observation of the activity of a large number of elements.

In the field of neuroscience, the multivariate time series typically obtained from brain recordings serve to infer minimal (effective) network models which can explain the dynamics of the nodes in a neural system. The motivation for such models can be, for instance, to describe a causal network \citep{Friston1994,Ay2008} or to model the directed information flow in the system \citep{Vicente2011} in order to produce a minimal computationally equivalent network \citep{Lizier2012a}.

Information theory \citep{Shannon1948, Cover2005} is well suited for the latter motivation of inferring networks that describe information flow as it provides model-free measures that can be applied at different scales and to different types of recordings. These measures, including conditional mutual information \citep{Cover2005} and transfer entropy \citep{Schreiber2000}, are based purely on probability distributions and are able to identify nonlinear relationships \citep{Palus1993}. Most importantly, information-theoretic measures allow the interpretation of the results from a distributed computation or information processing perspective, by modelling the information storage, transfer, and modification within the system \citep{Lizier2013}. Therefore, information theory simultaneously provides the tools for building the network model and the mathematical framework for its interpretation.

The general approach to network model construction can be outlined as follows: for any \emph{target} process (element) in the system, the inference algorithm selects the \emph{minimal set} of processes that collectively contribute to the computation of the target's next state. Every process can be separately studied as a target and the results can be combined into a directed network describing the information flows in the system.

This task presents several challenges:
\begin{itemize}
\item The state space of the possible network models grows faster than exponentially with respect to the size of the network;
\item Information-theoretic estimators suffer from the ``curse of dimensionality'' for large sets of variables \citep{Roulston1999, Paninski2003};
\item In a network setting, statistical significance testing requires multiple comparisons. This results in a high false positive rate (type I errors) without adequate family-wise error rate controls \citep{Dickhaus2014} or a high false negative rate (type II errors) with naive control procedures;
\item Nonparametric statistical testing based on shuffled surrogate time series is computationally demanding but currently necessary when using general information-theoretic estimators \citep{Lindner2011,Bossomaier2016}.
\end{itemize}

Several previous studies \citep{Vlachos2010, Faes2011, Lizier2012a, Sun2015} proposed greedy algorithms to tackle the first two challenges outlined above (see a summary by \citet[sec 7.2]{Bossomaier2016}). These algorithms mitigate the curse of dimensionality by greedily selecting the random variables that iteratively reduce the uncertainty about the present state of the target. The reduction of uncertainty is rigorously quantified by the information-theoretic measure of conditional mutual information (CMI), which can also be interpreted as a measure of conditional independence \citep{Cover2005}. In particular, these previous studies employed multivariate forms of the transfer entropy, \ie,~conditional and collective forms \citep{Lizier2008,Lizier2010}.
In general, such greedy optimisation algorithms provide a locally optimal solution to the NP-hard problem of selecting the most informative set of random variables. An alternative optimisation strategy---also based on conditional independence---employs a preliminary step to prune the set of sources \citep{Runge2012, Runge2018a}. Despite this progress, the computational challenges posed by the estimation of multivariate transfer entropy have severely limited the size of problems investigated in previous validation studies in the general case of nonlinear estimators, \eg,~\citet{Montalto2014} used \num{5} nodes and \num{512} samples; \citet{Kim2016} used \num{6} nodes and \num{100} samples; \citet{Runge2018a} used \num{10} nodes and \num{500} samples. However, modern neural recordings often provide hundreds of nodes and tens of thousands of samples.

These computational challenges, as well as the multiple testing challenges described above, are addressed here by the implementation of rigorous statistical tests, which represent the main theoretical contribution of this paper. These tests are used to control the family-wise error rate and are compatible with parallel processing, allowing the simultaneous analysis of the targets. This is a crucial feature, which enabled an improvement on the previous greedy algorithms. Exploiting the parallel computing capabilities of high-performance computing clusters and graphics processing units (GPUs) enabled the analysis of networks at a relevant scale for brain recordings---up to \num{100} nodes and \num{10000} samples. Our algorithm has been implemented in the recently released IDTxl Python package (the ``Information Dynamics Toolkit xl'' \citep{Wollstadt2019}).\ifarXiv\footnote{The ``Information Dynamics Toolkit xl'' is an open-source Python package available on GitHub (\url{https://github.com/pwollstadt/IDTxl}). In this paper, we refer to the current release at the time of writing (v1.0).}\else\jargon{IDTxl}{The ``Information Dynamics Toolkit xl'' is an open-source Python package available on GitHub (see Supportive Information).}\fi

We validated our method on synthetic datasets involving random structural networks of increasing size (also referred to as \emph{ground truth}) and different types of dynamics (vector autoregressive processes and coupled logistic maps). In general, effective networks are able to reflect dynamic changes in the regime of the system and do not reflect an underlying structural network. Nonetheless, in the absence of hidden nodes (and other assumptions, including stationarity and the causal Markov condition), the inferred information network was proven to reflect the underlying structure for a sufficiently large sample size \citep{Sun2015}.
Experiments under these conditions provide arguably the most important validation that the algorithm performs as expected, and here we perform the first large-scale empirical validation for non-Gaussian variables. As shown in the Results, the performance of our algorithm increased with the length of the time series, reaching consistently high precision, recall, and specificity ($>98\%$ on average) for \num{10000} time samples. Varying the statistical significance threshold showed a more favourable precision-recall trade-off for longer time series.

\section{Methods}
\subsection{Definitions and assumptions}
Let us consider a system of $N$ discrete-time stochastic processes for which a finite number of samples have been recorded (over time and/or in different replications of the same experiment).
In general, let us assume that the stochastic processes are stationary in each experimental time-window and Markovian with finite memory $l_\textup{M}$.\ifarXiv\footnote{The present state of the target does not depend on the past values of the target and the sources beyond a maximum finite lag $l_\textup{M}$.} \else\jargon{Markovian with finite memory}{The present state of the target does not depend on the past values of the target and the sources beyond a maximum finite lag $l_\textup{M}$.}\fi
Further assumptions will be made for the validation study.

\begin{figure}[htp]
\centering
\includegraphics[width=0.75\textwidth]{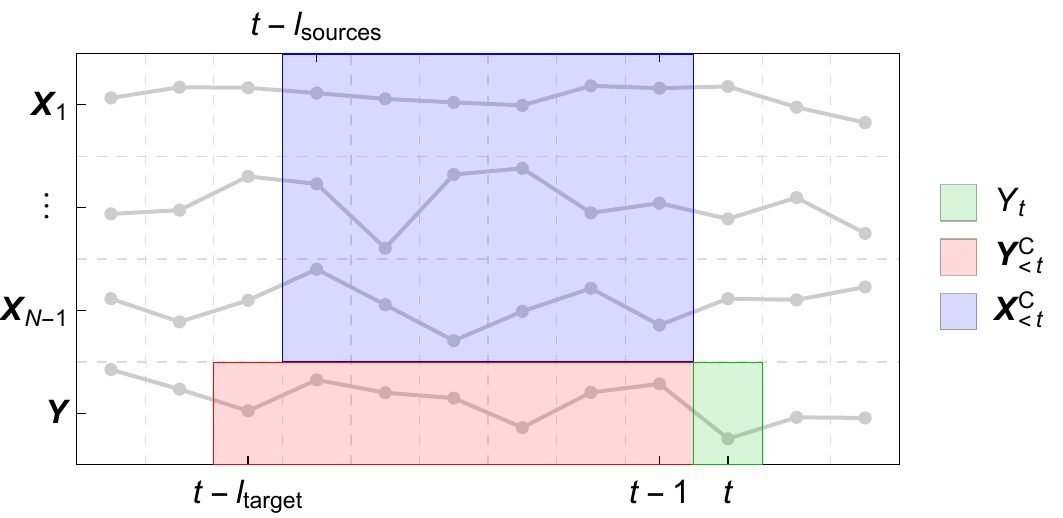}
\caption{Example of a possible definition of the candidate sets. The bottom row represents the time series of the target process $\bm{Y}$, with the present state $Y_t$ highlighted in green and the candidate target past set $\bm{Y}_{<t}^C$ highlighted in red (up to a lag $l_\textup{target}$). The remaining rows represent the time series of the source processes $\bm{X}_i$, with the candidate sources past set $\bm{X}_{<t}^C$ highlighted in blue (up to a lag $l_\textup{sources}$). For simplicity, only a single trial of the experiment is represented.}
\label{fig:candidate_sets}
\end{figure}

The following quantities are needed for the setup and formal treatment of the algorithm and are visualised in \fig{candidate_sets} and \fig{selected_sets}:
\begin{description}
\item [target process] $\bm{Y}$:
a process of interest within the system (where $\bm{Y}=\{Y_t \mid t \in \mathbb{N}\}$); the choice of the target process is arbitrary and all the processes in the system can separately be studied as targets;
\item [source processes] $\bm{X}_i$:
the remaining processes within the system (where $i=1,\ldots, N-1$ and $\bm{X}_i=\{X_{i,t} \mid t \in \mathbb{N}\}$);
\item [sample number (or size)] $T$:
the number of samples recorded over time;
\item [replication number] $R$:
the number of replications of the same experiment (\eg,~trials);
\item [target present state] $Y_t$:
the random variable (RV) representing the state of the target at time~$t$ (where $t \leq T$), whose information contributors will be inferred;
\item [candidate target past] $\bm{Y}_{<t}^C$:
an arbitrary finite set of RVs in the past of the target, up to a maximum lag $l_\textup{target}$, \ie,~$\bm{Y}_{<t}^C=\{Y_{t-1},\ldots,Y_{t-l_\textup{target}}\}$;
\item [candidate sources past] $\bm{X}_{<t}^C$:
an arbitrary finite set of RVs in the past of the sources, up to a maximum lag $l_\textup{sources}$, \ie,~$\bm{X}_{<t}^C=\{X_{i,t-1},\ldots,X_{i,t-l_\textup{sources}} \mid i=1,\ldots,N-1\}$;
\item [selected target past] $\bm{Y}_{<t}^S$:
the subset of RVs within the candidate target past set $\bm{Y}_{<t}^C$ that maximally reduces the uncertainty about the present state of the target;
\item [selected sources past] $\bm{X}_{<t}^S$:
the subset of RVs within the candidate sources past set $\bm{X}_{<t}^C$ that maximally further reduces the uncertainty about the present state of the target, in the context of the selected target past (explained in detail in the following section).
\end{description}

\subsection{Inference algorithm}
\label{sec:algorithm_steps}

For a given target process $\bm{Y}$, the goal of the algorithm is to infer the minimal set of information contributors to $Y_t$---defined as the selected sources past $\bm{X}_{<t}^S$---in the context of the relevant information contributors from the candidate target past set, defined as the selected target past $\bm{Y}_{<t}^S$.

\begin{figure}[htp]
\centering
\includegraphics[width=0.75\textwidth]{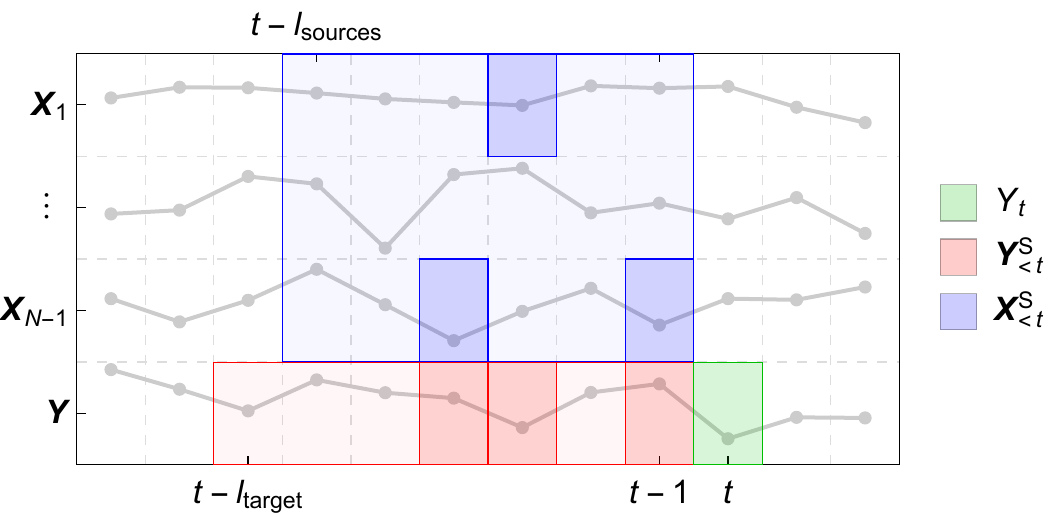}
\caption{Example of a resulting nonuniform embedding of the time series relevant to $Y_t$. The bottom row represents the time series of the target process $\bm{Y}$, with the present state $Y_t$ highlighted in green and the selected target past set $\bm{Y}_{<t}^S$ highlighted in red (as a subset of the candidate target past set shown in light red). The remaining rows represent the time series of the source processes $\bm{X}_i$, with the selected sources past set $\bm{X}_{<t}^S$ highlighted in blue (as a subset of the candidate sources past set shown in light blue). The embedding only specifies the \emph{relative} lags between the variables. For simplicity, only a single trial of the experiment is shown.}
\label{fig:selected_sets}
\end{figure}

The algorithm operates in four steps:
\begin{enumerate}
\item{Select variables in the candidate target past set $\bm{Y}_{<t}^C$ to obtain $\bm{Y}_{<t}^S$}
\item{Select variables in the candidate sources past set $\bm{X}_{<t}^C$ to obtain $\bm{X}_{<t}^S$}
\item{Prune the selected sources past variables}
\item{Test relevant variables collectively for statistical significance}
\end{enumerate}
\noindent The operations performed in the four steps are described in detail hereafter; the result is a \emph{nonuniform embedding} of the target and sources time series \citep{Takens1981, Vlachos2010,Faes2011}, as illustrated in \fig{selected_sets}.\ifarXiv\footnote{The term \emph{embedding} refers to the property of the selected set in capturing the underlying \emph{state} of the process as it relates to the target's next value, akin to a Takens' embedding \citep{Takens1981} yet with nonuniform delays between selected points \citep{Vlachos2010,Faes2011}.}\else\jargon{Nonuniform embedding}{A set of non-uniformly spaced time lags that captures the underlying \emph{state} of the process, akin to a Takens' embedding.}\fi

\ifarXiv
\subsubsection{Step 1: Select variables in the candidate target past set}
\else
\paragraph{Step 1: Select variables in the candidate target past set}
\fi

The goal of the first step is to find the subset of RVs within the candidate target past set $\bm{Y}_{<t}^C$ that maximally reduces the uncertainty about the present state of the target while meeting statistical significance requirements. Let $\bm{Y}_{<t}^S$ be the \emph{selected target past} set found via optimisation under these criteria.

Finding the globally optimal embedding is an NP-hard problem and requires testing all the subsets of the candidate target past set. Since the number of subsets grows exponentially with the size of the candidate set, this is computationally unfeasible; therefore, a greedy approximation algorithm is employed to find a locally optimal solution in the search space of possible embeddings. This approach tackles the challenge of computational complexity by aiming at identifying a minimal conditioning set; in doing so, it also tackles the curse of dimensionality in the estimation of information-theoretic functionals.

The set $\bm{Y}_{<t}^S$ is initialised as an empty set and it is iteratively built up via the following algorithm:
\begin{enumerate}[label=\alph*]
\item For each candidate variable $C \in \bm{Y}_{<t}^C$, estimate the CMI contribution $I(C;Y_t|\bm{Y}_{<t}^S)$;
\item Find the candidate $C^*$ which maximises the CMI contribution (reduction of uncertainty) and perform a statistical significance test against the null hypothesis of conditional independence, \ie,~that the new variable does not further reduce the uncertainty in the context of the previously included variables. If significant, add $C^*$ to $\bm{Y}_{<t}^S$ and remove it from $\bm{Y}_{<t}^C$. The \emph{maximum statistic} is employed to control the family-wise error rate (explained in detail in the \emph{Statistical tests} section);
\item Repeat the previous steps until the maximum CMI contribution is not significant or $\bm{Y}_{<t}^C$ is empty.
\end{enumerate}
\noindent From a distributed, intrinsic computation perspective, the goal can be interpreted as finding the embedding of the target's past states that maximises the \emph{active information storage}\ifarXiv\footnote{The active information storage is defined as the mutual information between the past and the present of the target: $I(\bm{Y}_{<t}^S;Y_t)$.} \else\jargon{Active information storage}{The mutual information between the past and the present of the target: $I(\bm{Y}_{<t}^S;Y_t)$.}\fi \citep{Lizier2012} to ensure self-prediction optimality as suggested by \citet{wibral2013}. This approach is similar to the one proposed by \citet{Garland2016} but uses nonuniform embedding and additional statistical controls.

The nonuniform embedding of the time series was introduced by \citet{Vlachos2010} and \citet{Faes2011}, who used an arbitrary threshold for the conditional mutual information. \citet{Lizier2012a} introduced a statistical significance test to select the candidates, which this study builds upon in proposing the maximum statistic. In addition, they embedded the target time series before embedding the sources, \ie,~the active information storage is modelled first and the information transfer is then examined in that context, thereby taking a specific modelling perspective on the information processing carried out by the system.

\ifarXiv
\subsubsection{Step 2: Select variables in the candidate sources past set}
\else
\paragraph{Step 2: Select variables in the candidate sources past set}
\fi

The goal of the second step is to find the subset of RVs within the candidate sources past set $\bm{X}_{<t}^C$ that maximally further reduces the uncertainty about the present state of the target, in the context of the selected target past found in the first step. Let $\bm{X}_{<t}^S$ be the \emph{selected sources past} set found via optimisation under these criteria.

As for step $1$, a greedy approximation algorithm is employed and the statistical significance is tested throughout the selection process. $\bm{X}_{<t}^S$ is initialised as an empty set and it is iteratively built up via the following algorithm:
\begin{enumerate}[label=\alph*]
\item For each candidate variable $C \in \bm{X}_{<t}^C$, estimate the conditional transfer entropy contribution $I(C;Y_t|\bm{Y}_{<t}^S,\bm{X}_{<t}^S)$ \citep{Verdes2005, Lizier2008, Lizier2010, Vakorin2009}. When $\bm{X}_{<t}^S$ is empty, this is simply a pairwise or bivariate transfer entropy \citep{Schreiber2000}; using the conditional form serves to prevent candidates carrying only redundant information (due to, \eg,~common driver or pathway effects) from being selected, as well as to capture synergistic interactions between $C$ and $\bm{X}_{<t}^S$.
\item Find the candidate $C^*$ which maximises the conditional transfer entropy contribution (reduction of uncertainty) and perform a statistical significance test against the null hypothesis of conditional independence: if significant, add $C^*$ to $\bm{X}_{<t}^S$ and remove it from $\bm{X}_{<t}^C$. The \emph{maximum statistic} is employed to control the family-wise error rate;
\item Repeat the previous steps until the maximum conditional transfer entropy contribution is not significant or $\bm{X}_{<t}^C$ is empty.
\end{enumerate}
\noindent From a distributed computation perspective, the goal can be interpreted as finding the nonuniform embedding of the source processes' past that maximises the \emph{collective transfer entropy} to the target, defined as $I(\bm{X}_{<t}^S;Y_t|\bm{Y}_{<t}^S)$ \citep{Lizier2010}. As above, the rationale for embedding the past of the sources as a second step is to achieve optimal separation of the storage and transfer contributions \citep{Lizier2012a}.

\ifarXiv
\subsubsection{Step 3: Prune the selected sources past variables}
\else
\paragraph{Step 3: Prune the selected sources past variables}
\fi

The third step of the algorithm is a pruning procedure performed to ensure that the variables included in the early iterations of the second step still provide a statistically-significant information contribution in the context of the final selected sources past set $\bm{X}_{<t}^S$. The pruning step involves the following operations:
\begin{enumerate}[label=\alph*]
\item For each variable $C \in \bm{X}_{<t}^S$, estimate the conditional mutual information contribution $I(C;Y_t|\bm{Y}_{<t}^S,\bm{X}_{<t}^S \setminus \{C\})$, where the the set difference operation is performed to exclude the variable $C$ from the conditioning set;
\item Find the variable $C^*$ which minimises the CMI contribution and perform a statistical significance test: if \emph{not} significant, remove $C$ from $\bm{X}_{<t}^S$. The \emph{minimum statistic} is employed to test for significance against the null hypothesis of conditional independence while controlling the family-wise error rate;
\item Repeat the previous steps until the minimum CMI contribution is not significant or $\bm{X}_{<t}^S$ is empty.
\end{enumerate}
\noindent The pruning step was introduced by \citet{Lizier2012a}; remarkably, \citet{Sun2015} proved that this step is essential for the theoretical convergence of the inferred network to the causal network in the Granger-Wiener framework; they also rigorously laid out the mathematical assumptions needed for such convergence (see \emph{Validation tasks} section).

\ifarXiv
\subsubsection{Step 4: Test relevant variables collectively for statistical significance}
\else
\paragraph{Step 4: Test relevant variables collectively for statistical significance}
\fi

The fourth and final step of the algorithm is the computation of the collective transfer entropy from the selected sources past set $\bm{X}_{<t}^S$ to the target and the performance of an \emph{omnibus test} to ensure statistical significance against the null hypothesis of conditional independence. The resulting omnibus p-value can further be used for correction of the family-wise error rate if the inference is carried out for multiple targets. The set $\bm{X}_{<t}^S$ is only accepted as a result if all the statistical tests are passed.
Importantly, the selected sources set $\bm{X}_{<t}^S$, inferred in the context of $\bm{Y}_{<t}^S$, is the final result of the algorithm for a given target process $\bm{Y}$. The order in which variables were inferred is not relevant.

The statistical tests play a fundamental role in the inference and provide the stopping conditions for the iterations involved in the first and second steps of the algorithm. These stopping conditions are adaptive and change according to the amount of data available (the length of the time series). Given their importance, the statistical tests are described in detail in the following section.

\subsection{Statistical tests}
\label{sec:statistical_tests}
The crucial steps in the inference algorithm rely on determining whether the CMI is positive. However, due to the finite sample size, the CMI estimators may produce non-zero estimates in the case of zero CMI and it may even return negative estimates if the estimator bias is larger than the true CMI \citep{Roulston1999, Kraskov2004}. For this reason, statistical tests are required to assess the significance of the CMI estimates against the null hypothesis of no CMI (\ie,~conditional independence) \citep{Chavez2003, Vicente2011, Lindner2011, Lizier2011}.

For certain estimators, analytic solutions exist for the finite-sample distribution under this null hypothesis (see \citet{Lizier2014JIDT}); in the absence of an analytic solution, the null distributions are computed in a nonparametric way by using surrogate time series \citep{Schreiber2000a}. The surrogates are generated to satisfy the null hypothesis by destroying the temporal relationship between the source and the target while preserving the temporal dependencies within the sources.

Finally, the inference algorithm is based on multiple comparisons and requires an appropriate calibration of the statistical tests to achieve the desired family-wise error rate (\ie,~the probability of making one or more false discoveries, or \emph{type I errors}, when performing multiple hypotheses tests). The maximum statistic and minimum statistic tests employed in this study were specifically conceived to tackle these challenges.

\subsubsection{Maximum statistic test}
\label{sec:max-stat}
The maximum statistic test is a step-down statistical test\ifarXiv\footnote{A test which proceeds from the smallest to the largest p-value. When the first non-significant p-value is found, all the larger p-values are also deemed not significant.} \else\jargon{Step-down statistical test}{A test which proceeds from the smallest to the largest p-value. When the first non-significant p-value is found, all the larger p-values are also deemed not significant.}\fi used to control the family-wise error rate when selecting the past variables for the target and sources embeddings, which involves multiple comparisons.

Let us first consider the first step of the main algorithm and assume that we have picked the single candidate variable $C^*$ (from the candidate target past set $\bm{Y}_{<t}^C$) which maximises the CMI contribution. The maximum statistic test mirrors this selection process by picking the maximum value among the surrogates. Specifically, let $I^*:=I(C^*;Y_t|\bm{Y}_{<t}^S)$ be the maximum contribution (\ie,~the maximum statistic); the following algorithm is used to test $I^*$ for statistical significance:
\begin{enumerate}[label=\alph*]
\item For each $C_j \in \bm{Y}_{<t}^C$, generate $S$ surrogates time series $C'_{j,1},\ldots,C'_{j,S}$ and compute the corresponding surrogate CMI values $I'_{j,1}=I(C'_{j,1};Y_t|\bm{Y}_{<t}^S),\ldots,I'_{j,S}=I(C'_{j,S};Y_t|\bm{Y}_{<t}^S)$. More details about the surrogate generation are provided at the end of this section. The number of surrogates $S$ must be chosen according to the desired significance level $\alpha_\textup{max}$, \ie,~such that $S>1/\alpha_\textup{max}$.
\item Compute the maximum CMI value over candidates $I^*_s:=max(I'_{1,s},\ldots,I'_{n,s})$ for each surrogate $s=1,\ldots,S$. Here, $n$ denotes the number of candidates and hence the number of comparisons. The obtained values $I^*_1,\ldots,I^*_S$ provide the (empirical) null distribution of the maximum statistic (see \tb{surrogates}).
\item Calculate the p-value for $I^*$ as the fraction of surrogate maximum statistic values that are larger than $I^*$.
\item $I^*$ is deemed \emph{significant} if the p-value is smaller than $\alpha_\textup{max}$ (\ie,~the null hypothesis of conditional independence for the candidate variable with the maximum CMI contribution is rejected at level $\alpha_\textup{max}$).
\end{enumerate}

\noindent The variables and quantities used in the above algorithm are presented in \tb{surrogates}.
The key goal in the surrogate generation is to preserve the temporal order of samples in the target time series $Y_t$ (which is not shuffled) and preserve the distribution of the sources $C_j$ while destroying any potential relationships between the sources and the target \citep{Vicente2011}. This can be achieved in multiple ways. If multiple replications (\eg,~trials) are available, surrogate data is generated by shuffling the order of replications for the candidate $C_j$ while keeping the order of replications for the remaining variables intact. When the number of replications is not sufficient to guarantee enough permutations, the embedded source samples within individual trials are shuffled instead (see \cite{Vicente2011, Lizier2011, Chavez2003, Verdes2005} and the summary by \citet[Appendix A.5]{Lizier2014JIDT}). Note that the generation of surrogates (steps a-c) can be avoided when the null distributions can be derived analytically, \eg,~with Gaussian estimators \citep{Barnett2012}.

The same test is performed during the selection of the variables in the candidate sources past set (step 2 of the main algorithm), with the only difference that $C_j \in \bm{X}_{<t}^C$ and that $\bm{X}_{<t}^S$ is added to the conditioning set, \ie~$I'_{j,s}=I(C'_{j,s};Y_t|\bm{Y}_{<t}^S, \bm{X}_{<t}^S)$ for each surrogate $s=1,\ldots,S$.

\begin{table}[htp]
\centering
\begin{tabular}{ccccccccccc}
\ifarXiv\hline\else\toprule\fi
 & Variable & CMI & \multicolumn{4}{c}{Surrogate variables} & \multicolumn{4}{c}{Surrogate CMI} \\
\ifarXiv\else\cmidrule(lr){4-7}\cmidrule(lr){8-11}\fi
& $C_j \in \bm{Y}_{<t}^C$ & $I_j=I(C_j;Y_t|\bm{Y}_{<t}^S)$ & $1$ & $2$ & $\cdots$ & $S$ & $1$ & $2$ & $\cdots$ & $S$\\
\ifarXiv\hline\else\midrule\fi
& $C_{1}$ & $I_{1}$ & $C'_{1,1}$ & $C'_{1,2}$ & $\cdots$ & $C'_{1,S}$ & $I'_{1,1}$ & $I'_{1,2}$ & $\cdots$ & $I'_{1,S}$ \\
& $C_{2}$ & $I_{2}$ & $C'_{2,1}$ & $C'_{2,2}$ & $\cdots$ & $C'_{2,S}$ & $I'_{2,1}$ & $I'_{2,2}$ & $\cdots$ & $I'_{2,S}$ \\
& $\vdots$ & $\vdots$ & $\vdots$ & $\vdots$ & & $\vdots$ & $\vdots$ & $\vdots$ & & $\vdots$ \\
& $C_{n}$ & $I_{n}$ & $C'_{n,1}$ & $C'_{n,2}$ & $\cdots$ & $C'_{n,S}$  & $I'_{n,1}$ & $I'_{n,2}$ & $\cdots$ & $I'_{n,S}$ \\
\ifarXiv\hline\else\midrule\fi
max CMI & & $I^*$  & & & & & $I^*_1$ & $I^*_2$ & $\cdots$ & $I^*_S$ \\
\ifarXiv\hline\else\bottomrule\fi
\end{tabular}
\caption{Computing the null distribution of the maximum statistic. The null distribution is empirically described by the values $I^*_1,\ldots,I^*_S$, obtained as $I^*_s:=max(I'_{1,s},\ldots,I'_{n,s})$, for each surrogate $s=1,\ldots,S$. Here, $n$ denotes the number of candidates and hence the number of comparisons. The null distribution is used to test the significance of $I^*$ against the null hypothesis of zero CMI.}
\label{tb:surrogates}
\end{table}

\subsubsection{Family-wise error rate correction}
\label{sec:FWER}
How does the maximum statistic test control the family-wise error rate? Intuitively, one or more statistics will exceed a given threshold if and only if the maximum exceeds it. This relationship can be used to obtain an adjusted threshold from the distribution of the maximum statistic under the null hypothesis, which can be used to control the family-wise error rate both in the weak and strong sense \citep{Nichols2003}.

Let us quantify the false positive rate $v_\textup{FPR}$ for a single variable when the maximum statistic at the significance level $\alpha_\textup{max}$ is employed.
For simplicity, the derivation is performed under the hypothesis that the information contributors to the target have been selected in the first iterations of the greedy algorithm and removed from the candidate sources past set $\bm{X}_{<t}^C$. Under this hypothesis, the target is conditionally independent of the remaining $n$ variables in $\bm{X}_{<t}^C$ given the selected source and target variables. Let $I_1,\ldots,I_{n}$ be the corresponding CMI estimates and let $I_{\textup{max}}:=\max(I_1,\ldots,I_{n})$ be the maximum statistic. As discussed above, the estimates might be positive even under the conditional independence hypothesis, due to finite-sample effects. Since the estimates are independently obtained from shuffled time series, they are treated as i.i.d. RVs.

Let $i_{\textup{threshold}}$ be the critical threshold corresponding to the given significance value $\alpha_\textup{max}$, \ie,~$i_{\textup{threshold}}:=\sup\{x\in \mathbb{R}|P(I_{\textup{max}}\geq x)=\alpha_{\textup{max}}\}$. Then
\begin{align}
	\alpha_{\textup{max}}&=P\left(I_{\textup{max}}> i_{\textup{threshold}}\right) \nn\\
	&=1-P\left(I_1\leq i_{\textup{threshold}},\ldots,I_{n}\leq i_{\textup{threshold}}\right) \nn\\
	&=1-\prod_{j=1}^{n} P\left(I_j\leq i_{\textup{threshold}}\right) \nn\\
	&=1-P\left(I_1\leq i_{\textup{threshold}}\right)^{n} \nn\\ 
	&=1-(1-v_\textup{FPR})^{n} 
\intertext{Therefore,}
	v_\textup{FPR}&=1-\left(1-\alpha_{\textup{max}}\right)^{1/n} \label{eq:FPR_variable}
\end{align}
Interestingly, equation \eq{FPR_variable} shows that the maximum statistic correction is equivalent to the Dunn-\v{S}id\'{a}k correction \citep{Sidak1967}.
Performing a Taylor expansion of \eq{FPR_variable} around $\alpha_{\text{max}}=0$ yields:
\begin{align} \label{eq:taylor_series}
	v_\textup{FPR}&=
	\sum_{j=1}^{\infty} \frac{-\prod\limits_{k=0}^{j-1} (k n-1)}{j!}\left(\frac{\alpha_{\text{target}}}{n}\right)^j
\end{align}
Truncating the Taylor series at $j=1$ yields the first-order approximation
\begin{equation}
	v_\textup{FPR}\approx \frac{\alpha_{\text{max}}}{n},
\end{equation}
which coincides with the false positive rate resulting from the Bonferroni correction \citep{Dickhaus2014}.
Moreover, since the summands in \eq{taylor_series} are positive for every $j$, the Taylor series is lower-bounded by any truncated series. In particular, the false positive rate resulting from the Bonferroni correction is a lower bound for the $v_\textup{FPR}$ (the false positive rate for a single variable resulting from the maximum statistic test), \ie,~the maximum statistic correction is less stringent than the Bonferroni correction.

Let us now study the effect of the maximum statistic test on the family-wise error rate $t_{\textup{FPR}}$ for a single target while accounting for all the iterations performed during the step-down test, (\ie,~$t_{\textup{FPR}}$ is the probability that at least one of the selected sources is a false positive). We have:

\begin{align} 
	t_{\textup{FPR}}&=\sum^{n}_{j=1} P(\text{``the source selected on step j is false positive''})\nn\\
	&=\sum^{n}_{j=1} \alpha_{\textup{max}}^j=\alpha_{\textup{max}}\left(\frac{1-\alpha_{\textup{max}}^n}{1-\alpha_{\textup{max}}}\right)
	\intertext{Therefore,}
	t_{\textup{FPR}} &\approx \alpha_{\textup{max}}
\end{align}
for the typical small values of $\alpha_{\textup{max}}$ used in statistical testing (even in the limit of large $n$), 
which shows that $\alpha_{\textup{max}}$ effectively controls the family-wise error rate for a single target.

\subsubsection{Minimum statistic test}
The minimum statistic test is employed during the third main step of the algorithm (pruning step) to remove the selected variables that have become redundant in the context of the final set of selected source past variables $\bm{X}_{<t}^S$, while controlling the family-wise error rate. This is necessary because of the multiple comparisons involved in the pruning procedure. The minimum statistic test works identically to the maximum statistic test (replacing ``maximum'' with ``minimum'' in the algorithm presented above).

\subsubsection{Omnibus test}
Let $T^*:=I(\bm{X}_{<t}^S;Y_t|\bm{Y}_{<t}^S)$ be the collective transfer entropy from all the selected sources past variables $\bm{X}_{<t}^S$ to the target $\bm{Y}$. The value $T^*$ is tested for statistical significance against the null hypothesis of zero transfer entropy (this test is referred to as the omnibus test). The null distribution is built using surrogates time series obtained via shuffling of the realisations of the selected sources (see \cite{Vicente2011, Lizier2011, Chavez2003, Verdes2005} and the summary by \citet[Appendix A.5]{Lizier2014JIDT}), \ie,~using a similar procedure to the one described in the \emph{Maximum statistic test} section above. Testing all the selected sources collectively is in line with the perspective that the goal of the network inference is to find the \emph{set} of relevant sources for each node.

\subsubsection{Combining across multiple targets}
When the inference is performed on multiple targets, the omnibus p-values can be employed in further statistical tests to control the family-wise error rate for the overall network (\eg,~via FDR-correction \citep{Benjamini1995, Dickhaus2014}, which is implemented in the IDTxl toolbox).

It is important to fully understand the statistical questions and validation procedure implied by this approach. Combining the results across multiple targets by reusing the omnibus test p-values for the FDR-correction yields a \emph{hierarchical} test. The test answers two nested questions: (1) \emph{'which nodes receive any significant overall information transfer?'} and, if any, (2)\emph{'what is the structure of the incoming information transfer to each node?'}. However, the answers are computed in the reverse order, for the following reason: it would be computationally unfeasible to directly compute the collective transfer entropy from all candidate sources to the target right at the beginning of the network inference process. At this point, the candidate source set usually contains a large number of variables so that estimation will likely fall prey to the curse of dimensionality. Instead, a conservative \emph{approximation} of the collective information transfer is obtained by considering only a subset of the potential sources, \ie,~those deemed significant by the maximum and minimum statistic tests described in the previous sections. Only if this approximation of the total information transfer is also deemed significant by the omnibus test (as well as by the FDR test at the network level), then the subset of significant sources for that target is interpreted \emph{post-hoc} as the local structure of the incoming information transfer. This way, the testing procedure exhibits a hierarchical structure: the omnibus test operates at the higher (global) level concerned with the collective information transfer, whereas the minimum and maximum tests operate at the lower (local) level of individual source-target connections.

Compared to a non-hierarchical analysis with a correction for multiple comparisons across all links (\eg,~by network-wide Bonferroni correction or the use of the maximum statistic across all potential links), the above strategy buys both statistical sensitivity ('recall') and the possibility to trivially parallelise computations across targets. The price to be paid is that a link with a relatively strong information transfer into a node with non-significant overall incoming information transfer may get pruned, while a link with relatively weaker information transfer into a node with significant overall incoming information transfer will prevail. This behaviour clearly differs from a correction for multiple comparisons across all links. Arguably, this difference is irrelevant in many practical cases, although it could become noticeable for networks with high average in-degree and relatively uniform information transfer across the links. The difference can be reduced by setting a conservative critical threshold for the lower-level greedy analysis.

\subsection{Validation tasks}
For the purpose of the validation study, the additional assumptions of \emph{causal sufficiency}\ifarXiv\footnote{Causal sufficiency: The set of observed variables includes all their common causes (or the unobserved common causes have constant values).} \else\jargon{Causal sufficiency}{The set of observed variables includes all their common causes (or the unobserved common causes have constant values).}\fi
and the \emph{causal Markov condition}\ifarXiv\footnote{Causal Markov condition: A variable X is independent of every other past variable conditional on all of its direct causes.} \else\jargon[48pt]{Causal Markov condition}{A variable X is independent of every other past variable conditional on all of its direct causes.}\fi{}
were made, such that the inferred network was expected to closely reflect the structural network for a sufficiently large sample size \citep{Sun2015}. Although this is not always the case, experiments under these conditions allow the evaluation of the performance of the algorithm with respect to an expected ground truth. An intuitive definition of these conditions is provided here, while the technical details are discussed at length in \cite{Spirtes1993}.
Moreover, the intrinsic stochastic nature of the processes makes purely synergistic and purely redundant interactions unlikely (and indeed vanishing for large sample size), thus satisfying the \emph{faithfulness} condition \citep{Spirtes1993}.

The complete network inference algorithm implemented in the IDTxl toolkit (release v1.0) was validated on multiple synthetic datasets, where both the structural connectivity and the dynamics were known. Given the general scope of the toolkit, two dynamical models of broad applicability were chosen: a vector autoregressive process (VAR) and a coupled logistic maps process (CLM); both models are widely used in computational neuroscience \citep{Zalesky2014, Rubinov2009, Valdes-Sosa2011}, macroeconomics \citep{Sims1980, Lorenz1993}, and chaos theory \citep{Strogatz2015}.

The primary goal was to quantify the scaling of the performance with respect to the size of the network and the length of the time series. Sparse directed random Erd\H{o}s-R\'{e}nyi networks \citep{Erdos1959} of increasing size ($N=$~\num{10} to \num{100} nodes) were generated with a link probability $p=3/N$ to obtain an expected in-degree of 3 links. Both the VAR and the CLM stochastic processes were repeatedly simulated on each causal network with increasingly longer time series ($T=$~\num{100} to \num{10000} samples), a single replication (or trial, \ie,~$R=1$), and with \num{10} random initial conditions. The performance was evaluated in terms of precision, recall, and specificity in the classification of the links. Further simulations were carried out to investigate the influence of the critical alpha level for statistical significance and the performance of different estimators of conditional mutual information.

\subsubsection{Vector autoregressive process}
\label{sec:VAR}
The specific VAR process used in this study is described by the following discrete-time recurrence relation:
\begin{equation}
Y_t=\beta Y_{t-1}+\sum_{\bm{X} \in \bm{X}_{\bm{Y}}} \alpha_{\bm{X}} X_{t-l_{\bm{X}}}+ \eta_t
\end{equation}
where $\bm{X}_{\bm{Y}}$ denotes the set of causal sources of the target process $\bm{Y}$ and a single random lag $l_{\bm{X}} \in \{1,2,3,4,5\}$ was used for each source $\bm{X}\in \bm{X}_{\bm{Y}}$. A Gaussian noise term $\eta_t$ with mean $\mu=0$ and standard deviation $\theta=0.1$ was added at each time step $t$; the noise terms added to different variables were uncorrelated.
The self-coupling coefficient was set to $\beta = 0.5$ and the cross-coupling coefficients $\alpha_{\bm{X}}$ were uniform and normalised for each target such that $\sum_{\bm{X} \in \bm{X}_{\bm{Y}}} \alpha_{\bm{X}} = 0.4$. This choice of parameters guaranteed that the VAR processes were stable (the resulting spectral radii were between \num{0.9} and \num{0.95}) and had stationary multivariate Gaussian distributions \citep{Atay2006}. As such, the Gaussian estimator implemented in IDTxl was employed for transfer entropy measurements in VAR processes.
Note that transfer entropy and Granger causality \citep{Granger1969} are equivalent for Gaussian variables \citep{Barnett2009}; therefore, using the Gaussian estimator with our algorithm can be viewed as extending Granger causality in the same multivariate/greedy fashion.

\subsubsection{Coupled logistic maps process}
\label{sec:CLM}
The coupled logistic maps process used in this study is described by the following discrete-time recurrence relations:
\begin{align}
a_t&=\beta Y_{t-1}+\sum_{\bm{X} \in \bm{X}_{\bm{Y}}} \alpha_{\bm{X}} X_{t-l_{\bm{X}}} \nn \\
Y_t&=(4 a_t(1-a_t)+ \eta_t) \ mod\ 1
\end{align}
At each time step $t$, each node $\bm{Y}$ computes the weighted input $a_t$ as a linear combination of its past value and the past of its sources, with the same conditions used for the VAR process on the choice of the random lags $l_{\bm{X}}$ and coupling coefficients $\beta$ and $\alpha_{\bm{X}}$. The value $Y_t$ is then computed by applying the logistic map activation function $f(x)=4x(1-x)$ to the weighted input $a_t$ and adding the Gaussian noise $\eta_t$ with the same properties used for the VAR process. Notice that the coefficient ($r=4$) used in the logistic map function corresponds to the fully-developed chaotic regime. The modulo-$1$ operation ensures that $Y_t\in [0,1]$ after the addition of noise. The \emph{nearest-neighbour} estimators were employed for transfer entropy measurements in the analysis of the CLM processes (in particular, Kraskov's estimator $I^{(1)}$ with $k=4$ nearest neighbours \citep{Kraskov2004} and its extension to CMI \citep{Frenzel2007, Vejmelka2008, Gomez-Herrero2015}). Nearest-neighbour estimators are model-free and are able to detect nonlinear dependencies in stochastic processes with non-Gaussian stationary distributions; fast CPU and GPU implementations are provided by the IDTxl package.

\section{Results}
\subsection{Influence of network size and length of the time series}
\label{sec:nodes_and_samples_sweep}
The aim of the first analysis was to quantify the scaling of the performance with respect to the size of the network and the length of the time series.

The inferred network was built by adding a directed link from a source node $\bm{X}$ to a target node $\bm{Y}$ whenever a significant transfer entropy from $\bm{X}$ to $\bm{Y}$ was measured while building the selected sources past set $\bm{X}_{<t}^S$ (\ie,~whenever $\bm{X} \cap \bm{X}_{<t}^S \neq \emptyset$). The critical alpha level for statistical significance was set to $\alpha_\textup{max}=0.001$ and $S = 1000$ surrogates were used for all experiments unless otherwise stated.
The candidate sets for the target as well as the sources were initialised with a maximum lag of five (\ie,~$l_\textup{target}=l_\textup{sources}=5$, corresponding to the largest lag values used in the definition of the VAR and CLM processes).

\begin{figure}[htp]
    \centering
    \begin{tabular}{cc}
        \centering
        \textbf{VAR}&\textbf{CLM}\\
    	\includegraphics[width=0.5\textwidth]{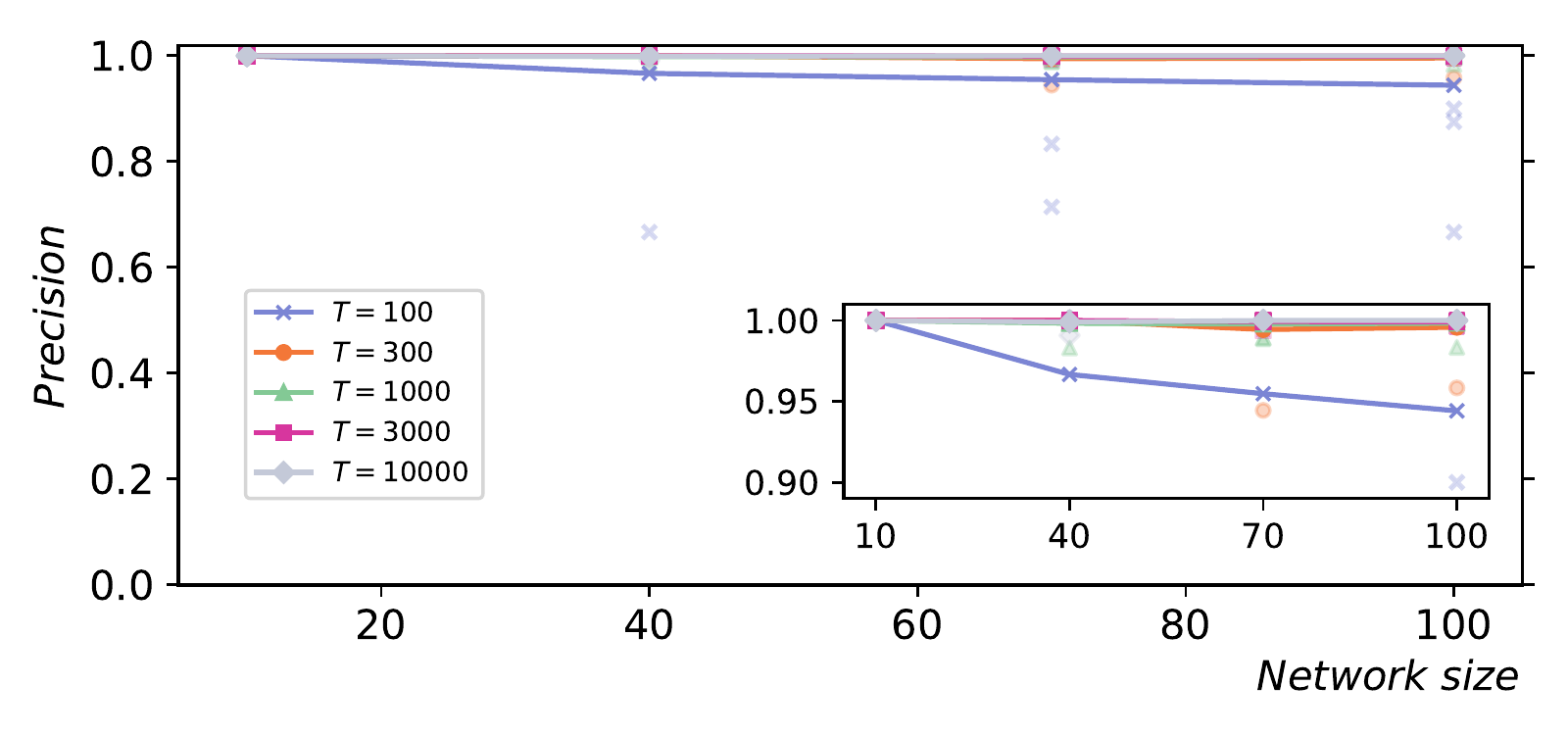}&\includegraphics[width=0.5\textwidth]{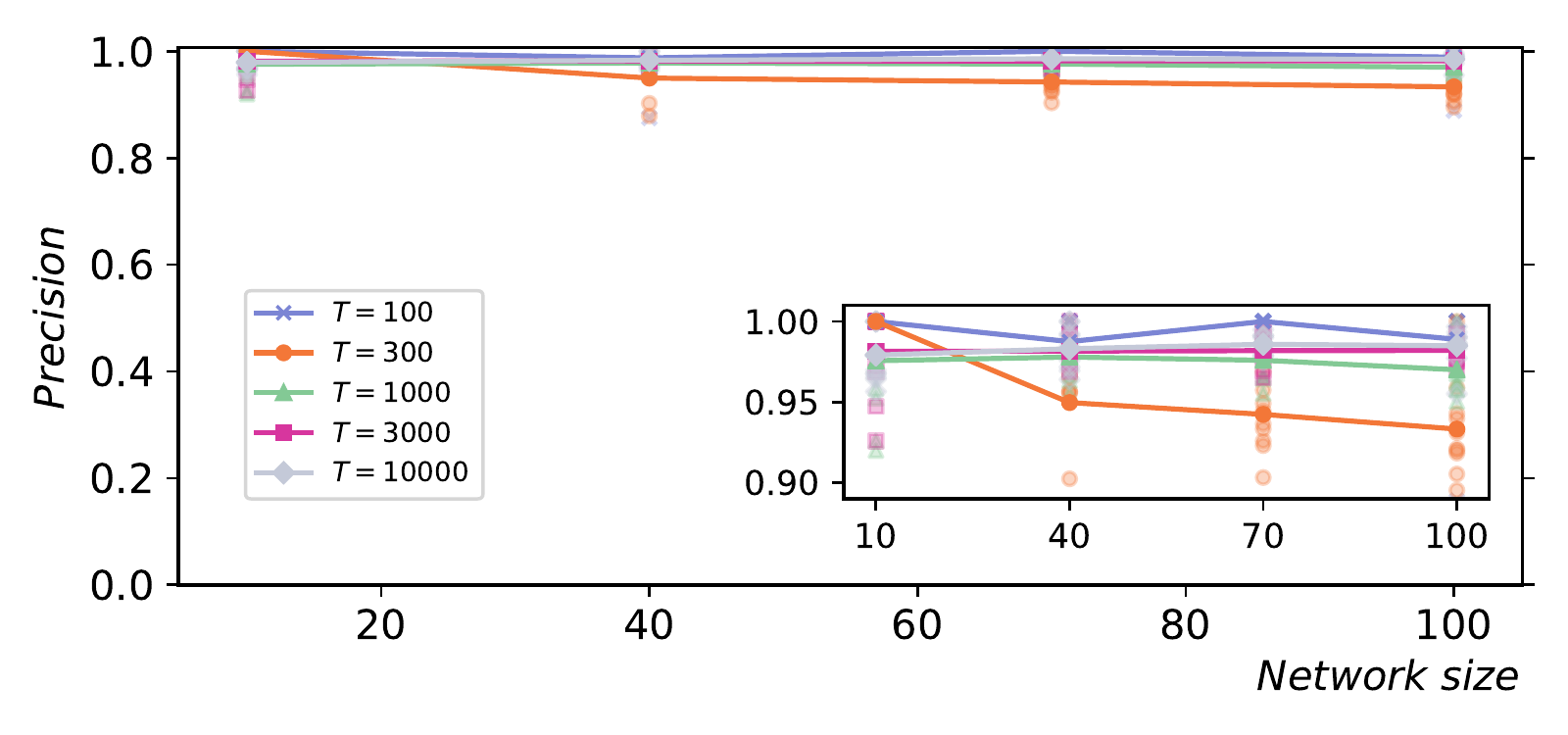}\\
    	\includegraphics[width=0.5\textwidth]{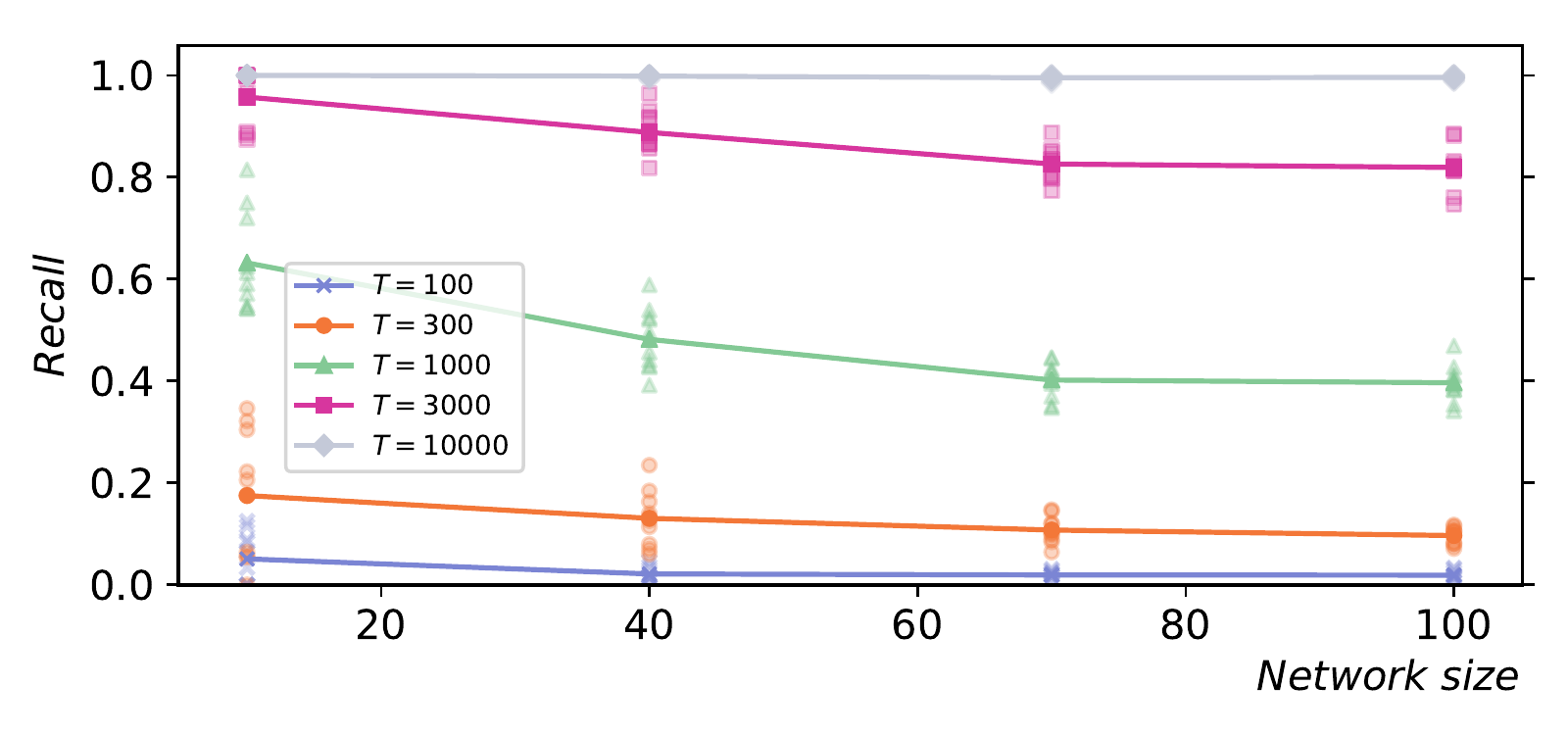}&\includegraphics[width=0.5\textwidth]{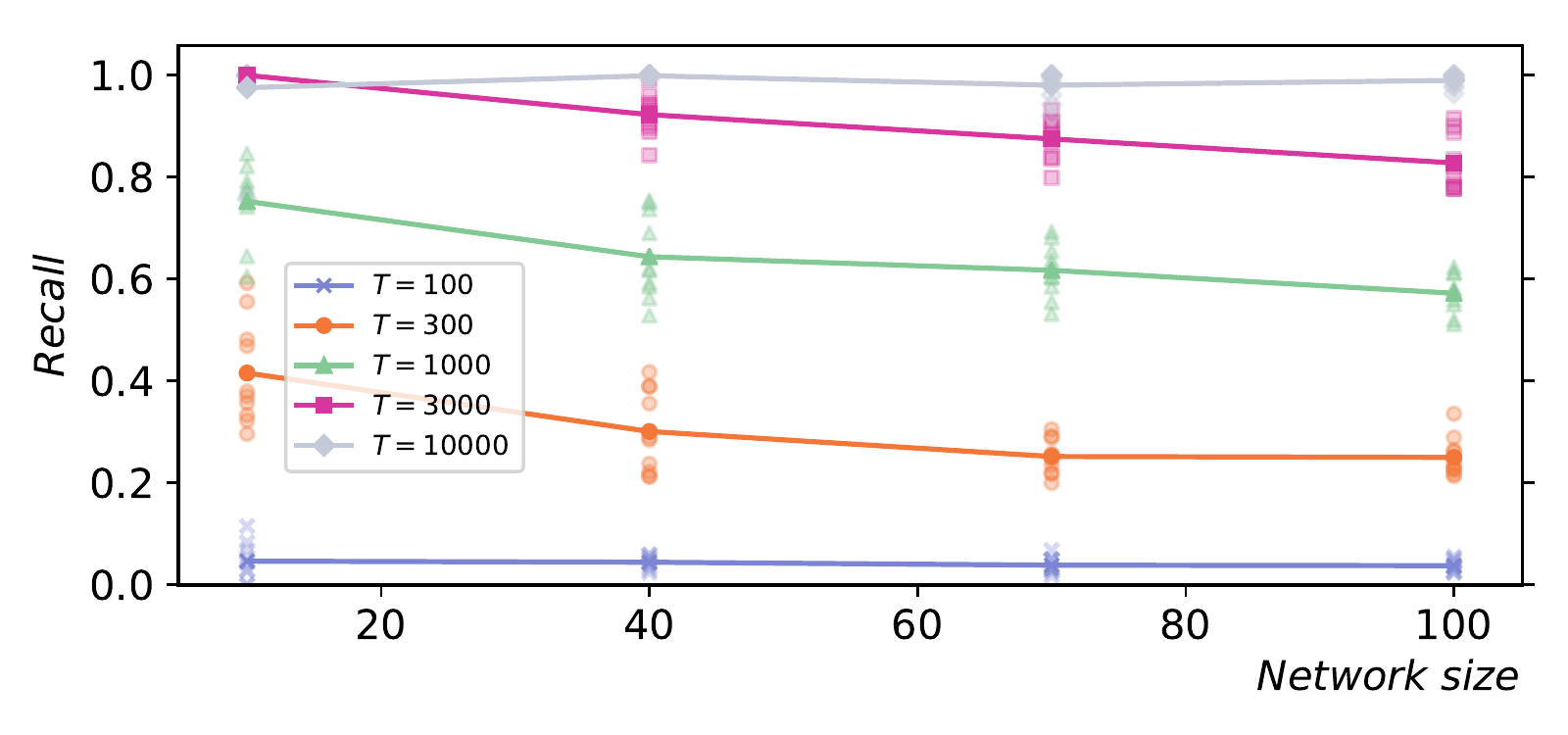}
	\end{tabular}
	\caption{Precision (top) and recall (bottom) for different network sizes, sample sizes, and dynamics. Left: Vector autoregressive process; Right: Coupled logistic maps. Each subplot shows five curves, corresponding to different time series lengths ($T=$~\num{100}, \num{300}, \num{1000}, \num{3000}, \num{10000}). The results for \num{10} simulations from different initial conditions are shown (low-opacity markers) in addition to the mean values (solid markers). All the random networks have an average in-degree $Np=3$.}
	\label{fig:nodes_and_samples_sweep}
\end{figure}

The network inference performance was evaluated in comparison to the known underlying structural network as a binary classification task, using standard statistics based on the number of \emph{true positives}~(TP, \ie,~correctly classified existing links), \emph{false positives}~(FP, \ie,~absent links falsely classified as existing), \emph{true negatives}~(TN, \ie,~correctly classified absent links), and \emph{false negatives}~(FN, \ie,~existing links falsely classified as absent). The following standard statistics were employed in the evaluation:
\begin{description}
\item[precision]
$=TP/(TP+FP)$
\item[recall]
$=TP/(TP+FN)$
\item[specificity]
$=TN/(TN+FP)$
\end{description}
The plots in \fig{nodes_and_samples_sweep} summarise the results in terms of precision and recall, while the specificity is additionally plotted in the Supplementary materials. For both types of dynamics, the performance increased with the number of samples and decreased with the size of the network.

For shorter time series ($T=$~\num{100} and $T=$~\num{1000}), the recall was the most affected performance measure as a function of $N$ and $T$, while the precision and the specificity were always close to optimal ($>98\%$ on average).
(Note that, while $S = 1000$ is minimal for $\alpha_\textup{max}=0.001$, recall was unchanged using $S=\num{10000}$ for $T=100$).
For longer time series ($T=$~\num{10000}), high performance according to all measures was achieved for both the VAR and CLM processes, regardless of the size of the network. The high precision and specificity are due to the effective control of the false positives, in accordance with the strict statistical significance level $\alpha_\textup{max}=0.001$ (the influence of $\alpha_\textup{max}$ is further discussed in the following sections). The inference algorithm was therefore conservative in the classification of the links.

\subsection{Validation of false positive rate}
The critical alpha level for statistical significance $\alpha_\textup{max}$ is a parameter of the algorithm which is designed to control the number o false positives in the network inference. As discussed in the \textit{Statistical tests} section in the Methods, $\alpha_\textup{max}$ controls the probability that a target is a false positive, \ie,~that at least one of its sources is a false positive. This approach is in line with the perspective that the goal of the network inference is to find the \emph{set} of relevant sources for each node.

\begin{figure}[htp]
	\centering
	\includegraphics[width=1\textwidth]{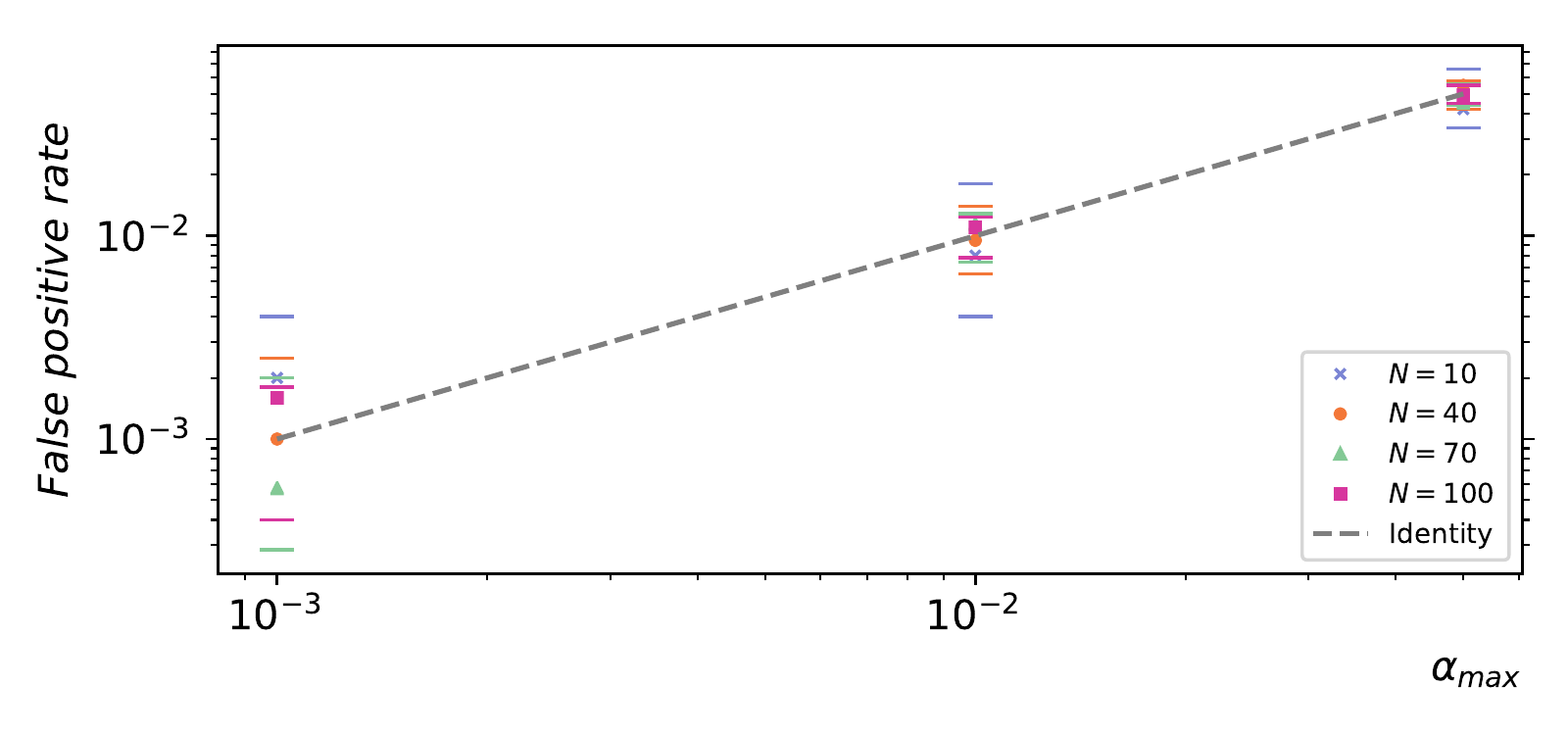}
	 \caption{Validation of false positive rate for a single target ($t_{\textup{FPR}}$) on empty networks. The points indicate the average false positive rate over \num{50} simulations of a vector autoregressive process ($T=$~\num{10000}). The horizontal marks indicate the corresponding $5$th and $95$th percentiles of the expected range.
	 These were computed empirically from the distribution of the random variable $\langle X_j/N \rangle$, where $X_j \sim Binomial(N, \alpha_\textup{max})$ are i.i.d. random variables, and the angular brackets indicate the finite average over $50$ repetitions. The $5$th percentile for $N=10$ and $N=40$ and $\alpha_\textup{max}=10^{-3}$ are equal to zero and therefore omitted from the log-log plot. The identity function is plotted as a reference (dashed line).}
	\label{fig:FPR_empty}
\end{figure}

A validation study was carried out to verify that the final number of false positives is consistent with the desired level $\alpha_\textup{max}$ after multiple statistical tests are performed. The \emph{false positive rate}\ifarXiv~(\ie~$FP/(FP+TN)$) \else\jargon{False positive rate}{$FP/(FP+TN)$}\fi
was computed after performing the inference on empty networks, where every inferred link is a false positive by definition (\ie,~under the complete null hypothesis). The rate was in good accordance with the critical alpha threshold $\alpha_\textup{max}$ for all network sizes, as shown in \fig{FPR_empty}.

The false positive rate validation was replicated in a scenario where the null hypothesis held for real fMRI data from the Human Connectome Project resting-state dataset (see Supporting Information). The findings are presented in the Supplementary Material, together with a note on autocorrelation. Notably, the results on fMRI data are in agreement with the results on synthetic data shown in \fig{FPR_empty}.

\subsection{Influence of critical level for statistical significance}
\label{sec:alpha_sweep_scatter}
Given the conservative results obtained for both the VAR and CLM processes (\fig{nodes_and_samples_sweep}), a natural question is to what extent the recall could be improved by increasing the critical alpha level $\alpha_\textup{max}$ and to what extent the precision would be negatively affected as a side effect.

\begin{figure}[htp]
	\centering
	\textbf{VAR}
	\\
	\includegraphics[width=1\textwidth]{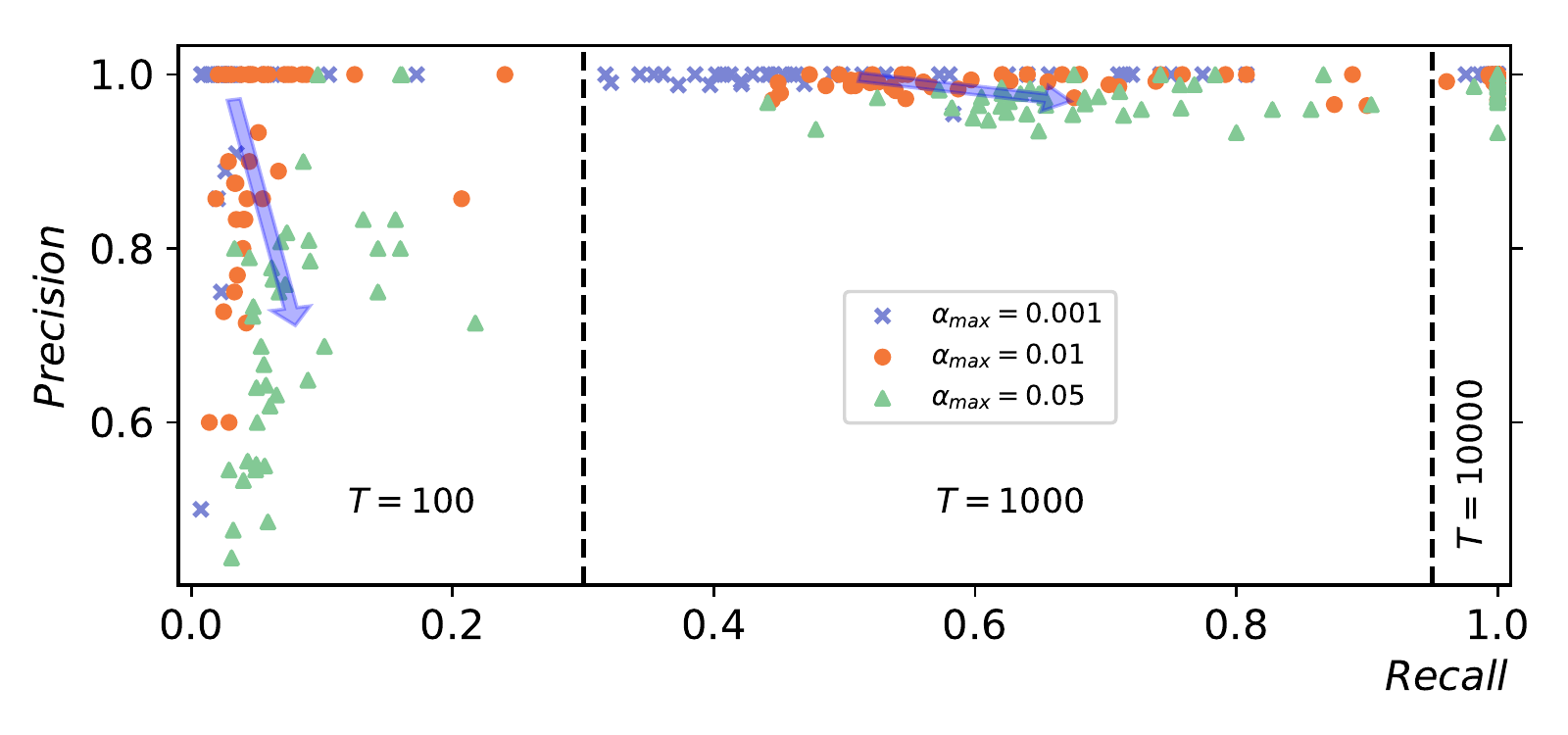}
	\\
	\textbf{CLM}
	\\
	\includegraphics[width=1\textwidth]{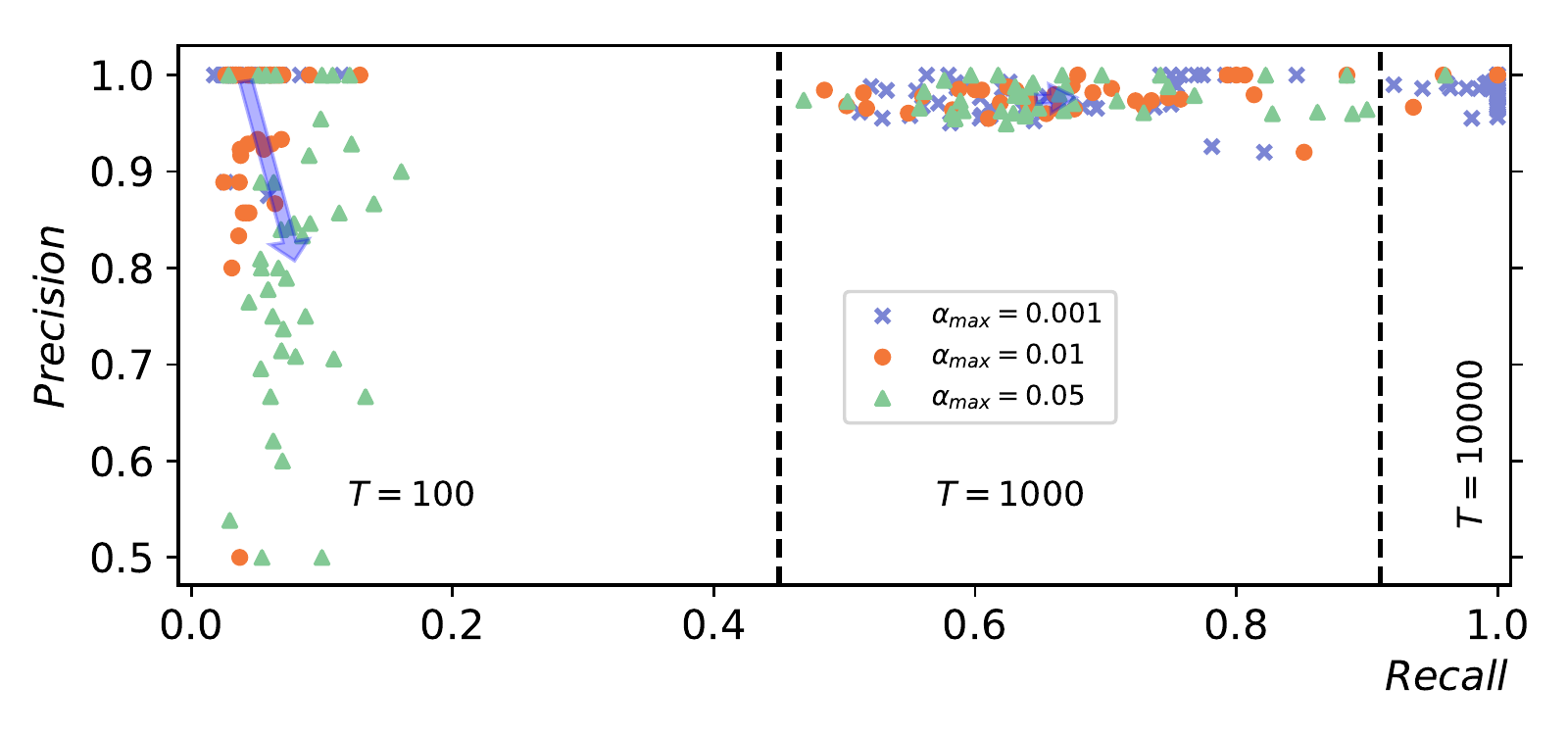}
	\caption{Influence of statistical significance threshold on network inference performance. Precision vs. recall for different statistical significance levels ($\alpha_\textup{max}=$ \num{0.05}, \num{0.01}, \num{0.001}), corresponding to different colours. The plots summarise the results for different dynamics (Top: Vector autoregressive process; Bottom: Coupled logistic maps), different time series lengths ($T=$~\num{100}, \num{1000}, \num{10000}), and different network sizes ($N=$~\num{10}, \num{40}, \num{70}, \num{100}, not distinguished). The arrows join the mean population values for the lowest and highest significance levels, illustrating the average trade-off between precision loss and recall gain.}
	\label{fig:alpha_sweep_scatter}
\end{figure}

In order to elucidate this trade-off, the analysis described above (\fig{nodes_and_samples_sweep}) was repeated for increasing values of $\alpha_\textup{max}$, with results shown in \fig{alpha_sweep_scatter}. For the shortest time series ($T=$~\num{100}), increasing $\alpha_\textup{max}$ resulted in a higher recall and a lower precision, as expected; on the other hand, for the longest time series ($T=$~\num{10000}), the performance measures were not significantly affected. Interestingly, for the intermediate case ($T=$~\num{1000}), increasing $\alpha_\textup{max}$ resulted in higher recall without negatively affecting the precision. 

\subsection{Inference of coupling lags}
So far, the performance evaluation focused on the identification the correct set of sources for each target node, regardless of the coupling lags. However, since the identification of the correct coupling lags is particularly relevant in neuroscience (see \citet{wibral2013} and references therein), the performance of the algorithm in identifying the correct coupling lags was additionally investigated.

\begin{figure}[htp]
	\centering
	\begin{tabular}{cc}
        \centering
        \textbf{VAR}&\textbf{CLM}\\
    	\includegraphics[width=0.5\textwidth]{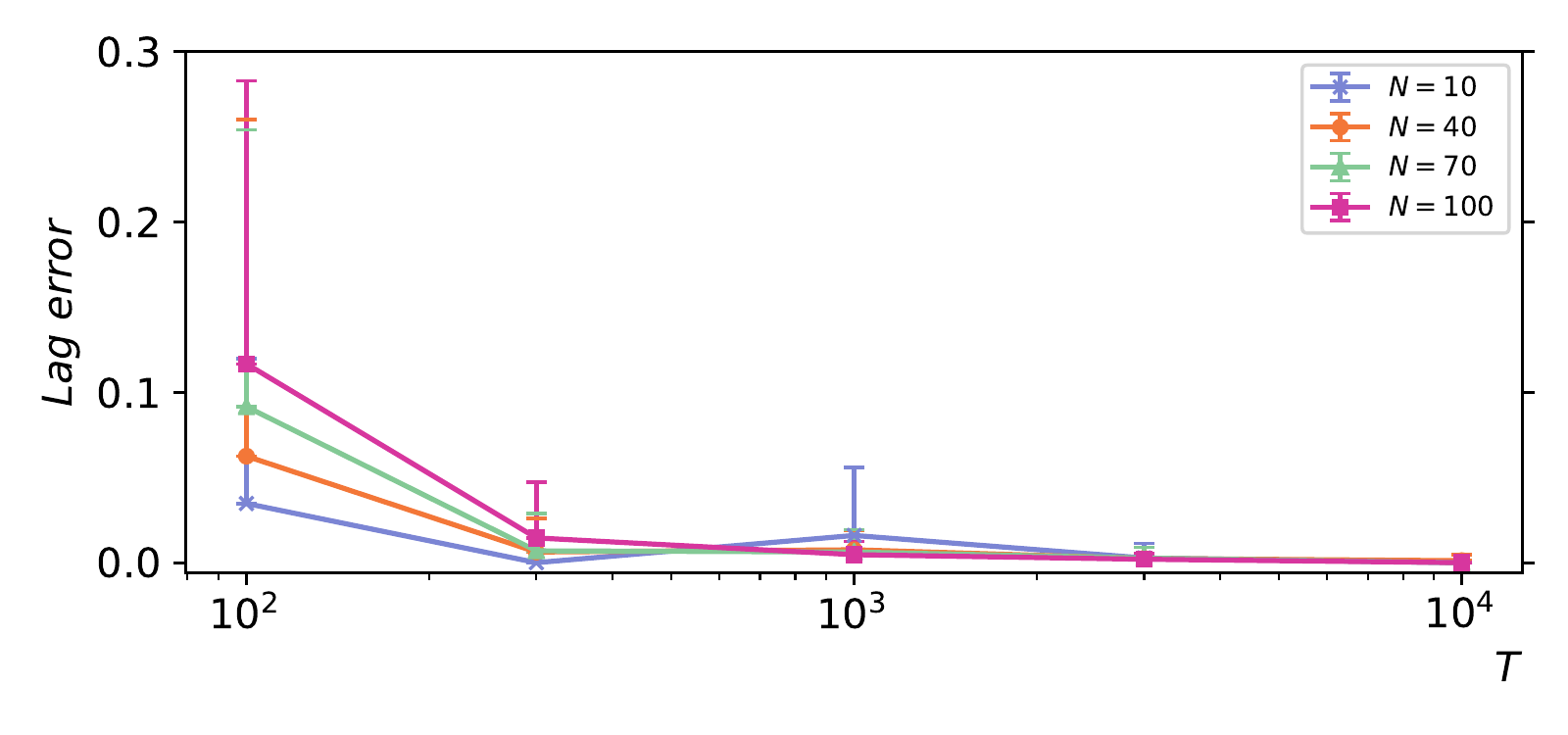}&\includegraphics[width=0.5\textwidth]{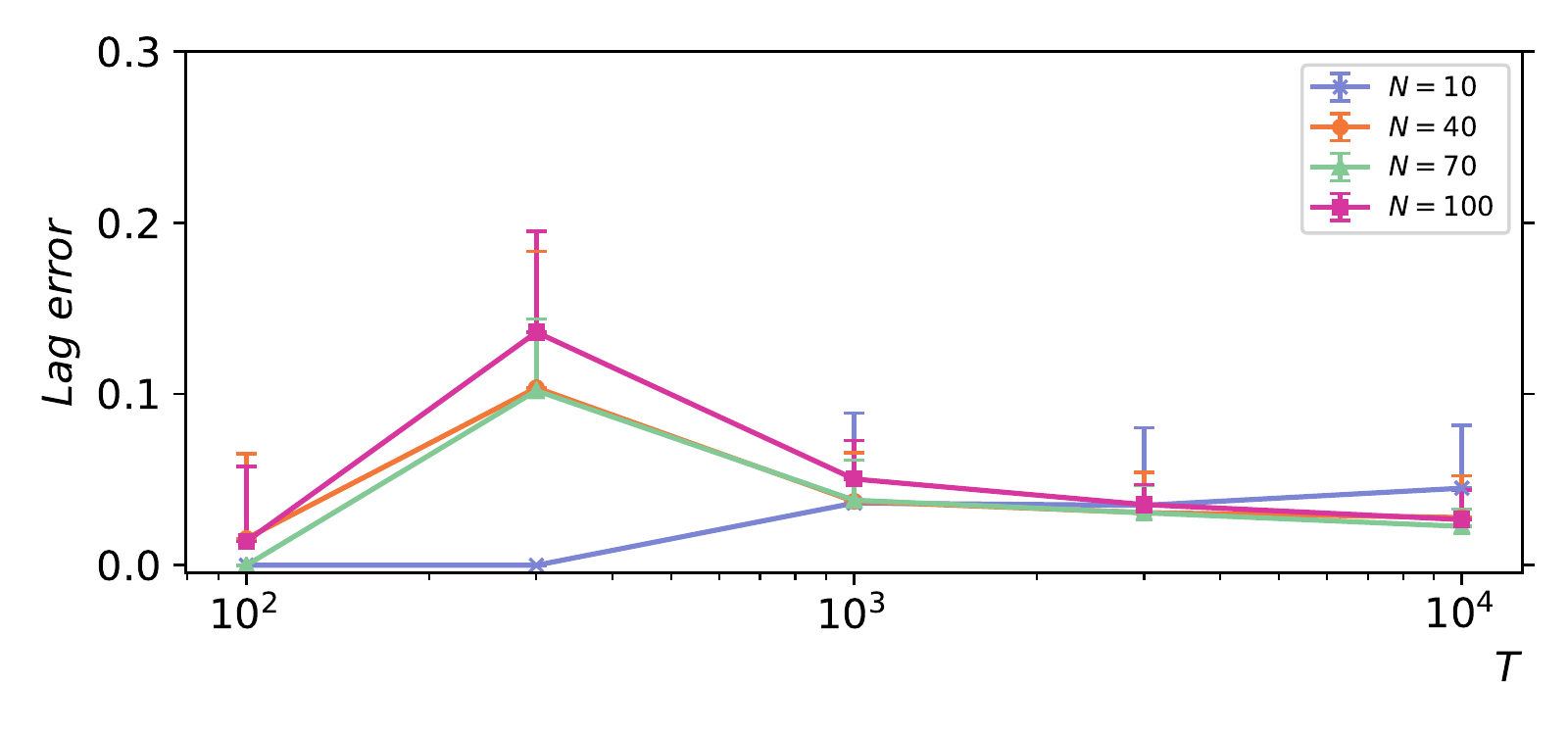}
	\end{tabular}
	\caption{Average absolute error between the real and the inferred coupling lags, relative to the value expected at random. Results for different dynamics (Left: Vector autoregressive process; Right: Coupled logistic maps), different time series lengths ($T=$~\num{100}, \num{300}, \num{1000}, \num{3000}, \num{10000}), and different network sizes ($N=$~\num{10}, \num{40}, \num{70}, \num{100}). The error bars indicate the standard deviation over \num{10} simulations from different initial conditions.}
	\label{fig:delay_error}
\end{figure}

By construction, a single coupling lag was imposed between each pair of processes (chosen at random between one and five discrete time steps, as described in the Methods). The average absolute error between the real and the inferred coupling lags was computed on the correctly recalled sources and divided by the value expected at random (which is the average absolute difference between two i.i.d. random integers in the $[1,5]$ interval). In line with the previous results on precision, the absolute error on coupling lag is consistently much smaller than that expected at random, even for the shortest time series (\fig{delay_error}). Furthermore, \num{1000} samples were sufficient to achieve nearly optimal performance for both the VAR and the CLM processes, regardless of the size of the network. Note that as $T$ increases and the recall increases, the lag error can increase (c.f. $T=100$ to $300$ for the CLM process). This is perhaps because while the larger $T$ permits more weakly contributing sources to be identified, it is not large enough to reduce the estimation error to make lag identification on these sources precise.

\subsection{Estimators}
\begin{figure}[htp]
	\centering
	\includegraphics[width=0.5\textwidth]{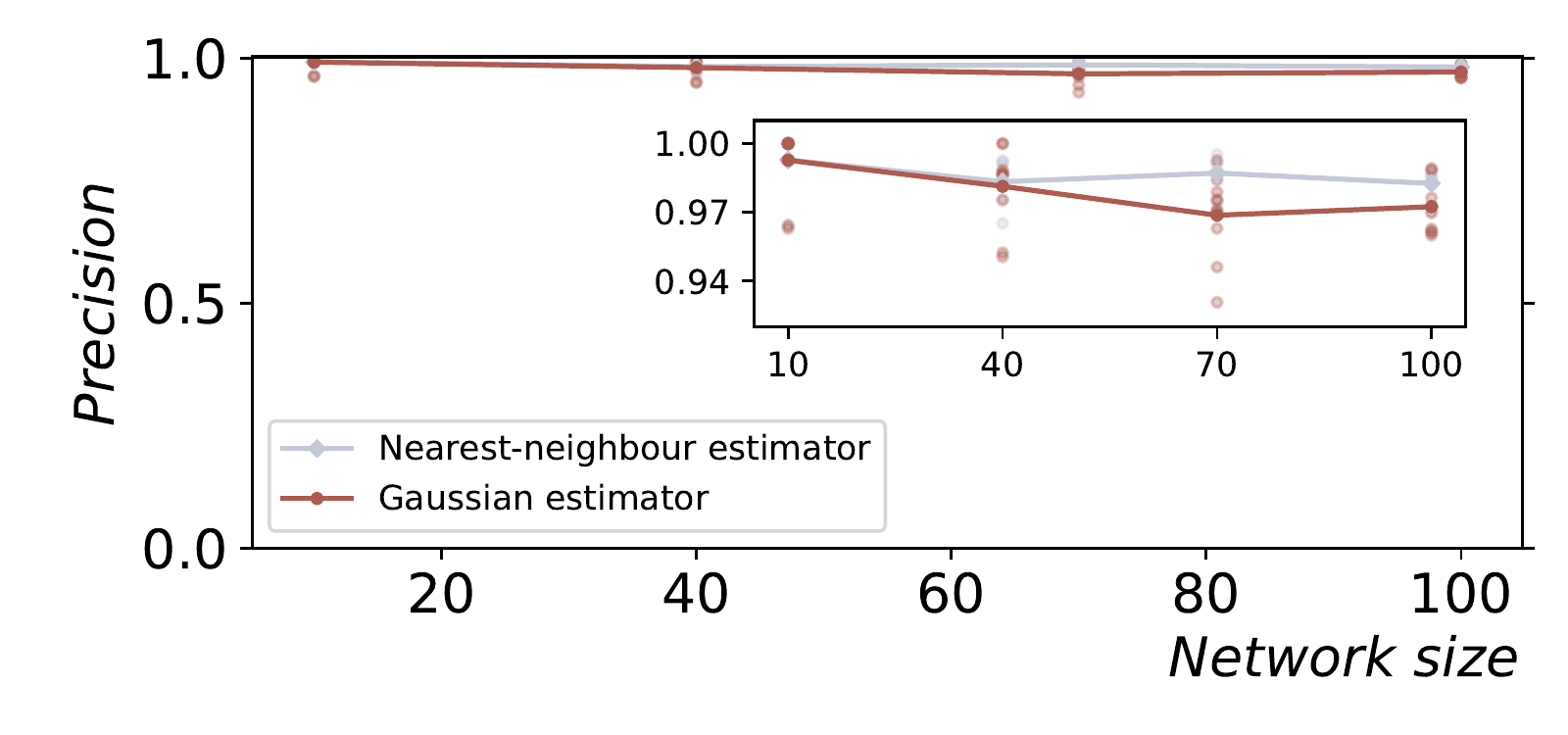}\includegraphics[width=0.5\textwidth]{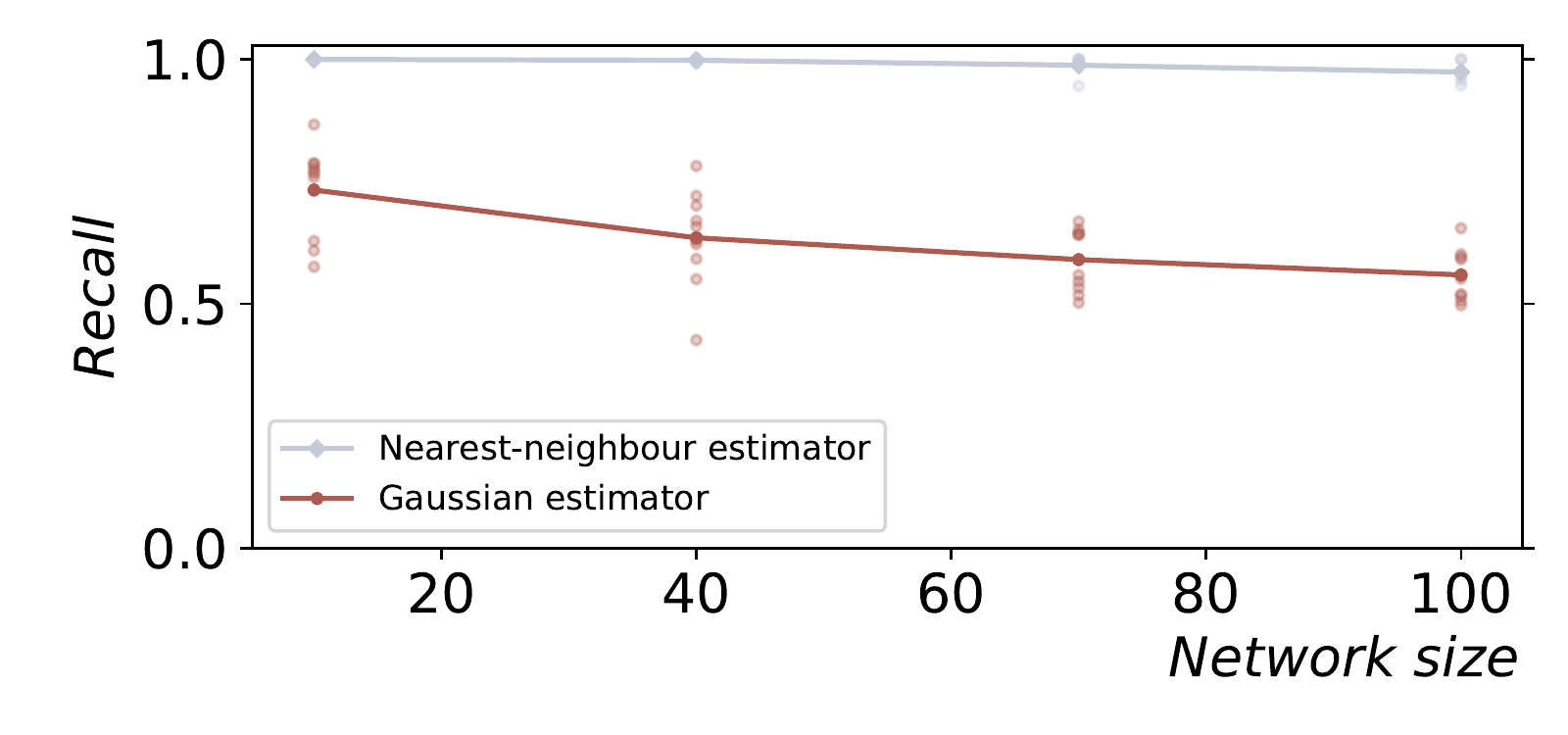}
	\caption{Gaussian vs. nearest-neighbour estimator on the coupled logistic maps process. The precision (left) and recall (right) are plotted against the network size and a fixed time series length ($T=$~\num{10000} samples). The results for \num{10} simulations from different initial conditions are shown (low-opacity markers) in addition to the mean values (solid markers). The statistical significance level $\alpha_\textup{max}=0.05$ was employed; an even larger gap between the recall of the estimators is obtained with $\alpha_\textup{max}=0.001$.}
	\label{fig:GC_vs_KSG_on_CLM}
\end{figure}

Given its speed, the Gaussian estimator is often used for large datasets or as a first exploratory step, even when the stationary distribution cannot be assumed to be Gaussian. The availability of the ground truth allowed to compare the performance of the Gaussian estimator and the nearest-neighbour estimator on the nonlinear CLM process, which does not satisfy the Gaussian assumption. As expected, the performance of the Gaussian estimator was lower than the performance of the nearest-neighbour estimator for all network sizes (\fig{GC_vs_KSG_on_CLM}).

The hierarchical tests introduced in the Methods section allow running the network inference algorithm in parallel on a high-performance computing cluster. Such parallelisation is especially needed when employing the nearest-neighbour estimator. In particular, each target node can be analysed in parallel on a CPU (employing one or more cores) or a GPU, which is made possible by the CPU and GPU estimators provided by the IDTxl package (custom OpenCL kernels were written for the GPU implementation). A summary of the CPU and GPU run-times is provided in the Supplementary materials.

\section{Discussion}
\label{sec:discussion}

The algorithm presented in this paper provides robust statistical tests for network inference to control the false positive rate. These tests are compatible with parallel computation on high-performance computing clusters, which enabled the validation study on synthetic sparse networks of increasing size (\num{10} to \num{100} nodes), using different dynamics (linear autoregressive processes and nonlinear coupled logistic maps) and increasingly longer time series (\num{100} to \num{10000} samples). Both the network size and the sample size are one order of magnitude larger than previously demonstrated, showing feasibility for typical EEG and MEG experiments. The results demonstrate that the statistical tests achieve the desired false positive rate and successfully address the multiple-comparison problems inherent in network inference tasks (\fig{FPR_empty}).

The ability to control the false positives while building connectomes is a crucial prerequisite for the application of complex network measures, to the extent that \cite{Zalesky2016} concluded that ``specificity is at least twice as important as sensitivity [\ie,~recall] when estimating key properties of brain networks, including topological measures of network clustering, network efficiency and network modularity''. The reason is that false positives occur more prevalently between network modules than within them and the spurious inter-modular connections have a dramatic impact on network topology \citep{Zalesky2016}.

The trade-off between precision and recall when relaxing the statistical significance threshold was further investigated (\fig{alpha_sweep_scatter}). When only \num{100} samples were used, the average recall gain was more than five times smaller than the average precision loss. In our opinion, this result is possibly due to the sparsity of the networks used in this study and suggests a conservative choice of the threshold for sparse networks and short time series. The trade-off was reversed for longer time series: when \num{1000} samples were used, the average recall gain was more than five times larger than the average precision loss. Finally, for \num{10000} samples, high precision and recall were achieved ($>98\%$ on average) for both the vector autoregressive and the coupled logistic maps processes, regardless of the statistical significance threshold.

For both types of dynamics, the network inference performance increased with the length of the time series and decreased with the size of the network (\fig{nodes_and_samples_sweep}). This is to be expected since larger systems require more statistical tests and hence stricter conditions to control the family-wise error rate (false positives). Specifically, larger networks result in wider null distributions of the maximum statistic (\ie,~larger variance), whereas longer time series have the opposite effect. Therefore, for large networks and short time series, controlling the false positives can have a negative impact on the ability to identify the true positives, particularly when the effect size (\ie,~the transfer entropy value) is small.

In addition, the superior ability of the nearest-neighbour estimator over the Gaussian estimator in detecting nonlinear dependencies was quantified. There is a critical motivation for this comparison: the general applicability of the nearest-neighbour estimators comes at the price of higher computational complexity and a significantly longer run-time, so that the Gaussian estimator is often used for large datasets (or at least as a first exploratory step), even when the Gaussian hypothesis is not justified. To investigate such scenario, the Gaussian estimator was tested on the nonlinear logistic map processes: while the resulting recall was significantly lower than the nearest-neighbour estimator for all network sizes, it was nonetheless able to identify over half of the links for a sufficiently large number (\num{10000}) of time samples (\fig{GC_vs_KSG_on_CLM}).

The stationarity assumption about the time series corresponds to assuming a single regime of neuronal activity in real brain recordings. If multiple regimes are recorded, which is typical in experimental settings (\eg,~sequences of tasks or repeated presentation of stimuli interleaved with resting time windows), different stationary regimes can be studied by performing the analysis within each time window. The networks obtained in different time windows can either be studied separately and compared against each other or collectively interpreted as a single evolving temporal network. To obtain a sufficient amount of observations per window, multiple replications of the experiment under the same conditions are typically carried out. Replications can be assumed to be cyclo-stationary and estimation techniques exploiting this property have been proposed \citep{Gomez-Herrero2015, Wollstadt2014}; these estimators are also available in the IDTxl Python package.
The convergence to the (unknown) causal network was only proven under the hypotheses of stationarity, causal sufficiency, and the causal Markov condition \citep{Sun2015}. However, conditional independence holds under milder assumptions \citep{Runge2018} and the absence of links is valid under general conditions. The conditional independence relationships can, therefore, be used to exclude variables in following intervention-based causal experiments, making network inference methods valuable for exploratory studies.

In fact, the directed network is only one part of the model and provides the scaffold over which the information-theoretic measures are computed. Therefore, even if the structure of a system is known and there is no need for network inference, information theory can still provide nontrivial insights on the distributed computation by modelling the information storage, transfer, and modification within the system \citep{Lizier2013}. This decomposition of the predictive information into the active information storage and transfer entropy components is one out of many alternatives within the framework proposed by \citet{Chicharro2012}. Arguably, the storage-transfer decomposition reflects the segregation-integration dichotomy that has long characterised the interpretation of brain function \citep{Zeki1988, Sporns2010}. Information theory has the potential to provide a quantitative definition of these fundamental but still unsettled concepts \citep{Li2019}. In addition, information theory provides a new way of testing fundamental computational theories in neuroscience, \eg,~predictive coding \citep{Brodski-Guerniero2017}.

As such, information-theoretic methods should not be seen as opposed to model-based approaches, but complementary to them \citep{Friston2013}. If certain physically motivated parametric models are assumed, the two approaches are equivalent for network inference: maximising the log-likelihood is asymptotically equivalent to maximising the transfer entropy \citep{Barnett2012, Cliff2018}. Moreover, different approaches can be combined, \eg,~the recent large-scale application of spectral DCM was made possible by using functional connectivity models to place prior constraints on the parameter space \citep{Razi2017}. Networks inferred using bivariate transfer entropy have also been employed to reduce the model space prior to DCM analysis \citep{Chan2017}.

In conclusion, the continuous evolution and combination of methods show that network inference from time series is an active field of research and there is a current trend of larger validation studies, statistical significance improvements, and reduction of computational complexity. Information-theoretic approaches require efficient tools to employ nearest-neighbour estimators on large datasets of continuous-valued time series, which are ubiquitous in large-scale brain recordings (calcium imaging, EEG, MEG, fMRI). The algorithm presented in this paper is compatible with parallel computation on high-performance computing clusters, which enabled the study of synthetic nonlinear systems of \num{100} nodes and \num{10000} samples. Both the network size and the sample size are one order of magnitude larger than previously demonstrated, bringing typical EEG and MEG experiments into scope for future information-theoretic network inference studies.
Furthermore, the statistical tests presented in the Methods are generic and compatible with any underlying conditional mutual information or transfer entropy estimators, meaning that estimators applicable to spike trains \citep{Spinney2017} can be used with this algorithm in future studies.


\section{Supporting Information}
The network inference algorithm described in this paper is implemented in the open-source Python software package IDTxl \citep{Wollstadt2019}, which is freely available on GitHub (\url{https://github.com/pwollstadt/IDTxl}). In this paper, we refer to the current release (v1.0) at the time of writing (\href{https://doi.org/10.5281/zenodo.2554339}{doi:10.5281/zenodo.2554339}).

The raw data used for the experiment presented in the Supplementary Material is openly available on the MGH-USC Human Connectome Project database (\url{https://ida.loni.usc.edu/login.jsp}).

\ifarXiv
\section{Acknowledgements}
\else
\acknowledgments
\fi
This research was supported by: Universities Australia/German Academic Exchange Service (DAAD) Australia-Germany Joint Research Cooperation Scheme grant: ``Measuring neural information synthesis and its impairment'', Grant/Award Number: PPP Australia Projekt-ID 57216857; Deutsche Forschungsgemeinschaft (DFG) Grant CRC 1193 C04; and Australian Research Council DECRA grant DE160100630 and Discovery grant DP160102742.

Data collection and sharing for the experiment presented in the Supplementary material was provided by the MGH-USC Human Connectome Project (HCP; Principal Investigators: Bruce Rosen, M.D., Ph.D., Arthur W. Toga, Ph.D., Van J. Weeden, MD). HCP funding was provided by the National Institute of Dental and Craniofacial Research (NIDCR), the National Institute of Mental Health (NIMH), and the National Institute of Neurological Disorders and Stroke (NINDS). HCP data are disseminated by the Laboratory of Neuro Imaging at the University of Southern California.

The authors acknowledge the Sydney Informatics Hub and the University of Sydney's high-performance computing cluster Artemis for providing the high-performance computing resources that have contributed to the research results reported within this paper. Furthermore, the authors thank Aaron J. Gutknecht for commenting on a draft of this paper, and Oliver Cliff for useful discussions and comments.

\ifarXiv
\section{Author Contributions}
\else
\authorcontributions
\fi

Leonardo Novelli: Conceptualization; Data curation; Formal analysis; Investigation; Software; Validation; Visualization; Writing --- original draft. Patricia Wollstadt: Conceptualization; Software; Writing --- review \& editing. Pedro Mediano: Software; Writing --- review \& editing. Michael Wibral: Conceptualization; Funding acquisition; Methodology; Software; Supervision; Writing --- review \& editing. Joseph T. Lizier: Conceptualization; Funding acquisition; Methodology; Software; Supervision; Writing --- review \& editing.

\bibliography{references}

\pagebreak

\ifarXiv
\else
\fi

\section{Supporting Information}

\subsection{Run-time on CPU and GPU}
\label{sec:run-time}

The number of transfer entropy calculations scales as $\mathcal{O}(N^2 d l_\textup{max} S)$, where $N$ is the number of processes, $d$ is the average inferred in-degree, $l_\textup{max}$ is the maximum temporal search depth per process (\ie,~$l_\textup{max}=max\{l_\textup{target}, l_\textup{sources}\}$), and $S$ is the number of surrogates. This assumes that $d$ is independent of the network size $N$; however, note that $d=N$ in the worst case of a fully connected network, leading to cubic run-times.

The network inference algorithm was designed for parallelisation:
\begin{itemize}
\item When using the CPU estimators, it is possible to parallelise over targets, resulting in $\mathcal{O}(N d l_\textup{max} S)$ transfer entropy calculations per target for the nearest-neighbour estimator and $\mathcal{O}(N d l_\textup{max})$ transfer entropy calculations for the Gaussian estimator (if using analytic null distributions instead of surrogates). The complexity of each calculation is $\mathcal{O}(k T log T)$ for the nearest-neighbour estimator and $\mathcal{O}(T)$ for the Gaussian estimator (where $T$ is number of time series samples and $k$ is the number of nearest-neighbours).
\item When using the GPU estimators, it is possible to parallelise both over targets and surrogates. Each target requires $\mathcal{O}(N d l_\textup{max})$ transfer entropy calculations including surrogates, assuming that all surrogates for a source fit into the GPU's main memory and can be processed in parallel. The complexity of each calculation is $\mathcal{O}(T^2)$. If enough memory is available, it is further possible to parallelise over time samples $T$, resulting in faster run-times in practice.
\end{itemize}

The practical run-time for a full network analysis that considers each process as a target depends on the number of available computing nodes. In the worst case, where only a single computing node is available, the full run-time is equal to the single-target run-time multiplied by $N$, since the target are analysed in series. In the best case, if $N$ computing nodes are available, the full run-time is equal to the single-target run-time, since all targets can be analysed in parallel.
Notice that there is a trade-off between run-time and memory requirements: if all the targets are analysed in parallel, the full required memory in $N$ times larger than the memory required in the single-node case; conversely, if the targets are analysed in series, the full required memory is equal to the memory required in the single-node case.

In the experiments presented in this article, the algorithm was either run using a \emph{single core} per target (on different Intel Xeon CPUs with similar characteristics: \num{2.1}-\num{2.6} GHz), or using a whole dedicated GPU per target (NVIDIA V100 SXM2, \num{16} GB RAM). These computations were performed on the Artemis computing cluster made available by the Sydney Informatics Hub at The University of Sydney. The maximum CPU and GPU run-times for a single target are shown in \tb{run-time}, which summarises the results for different time series lengths and different network sizes. Notice that the CPU run-time per target can be reduced if multiple cores per target are available.

\begin{table}[htp]
\centering
\begin{tabular}{cSSSS}
\ifarXiv\hline\else\toprule\fi
{Sample size} & {Network size} & {max GPU time} & {max CPU time} & {max CPU memory}\\
{} & {} & {per target} & {per target} & {per target}\\
{$T$} & {$N$} & {(h)} & {(h)} & {(GB)} \\
\ifarXiv\hline\else\midrule\fi
\multirow{4}*{\num{100}}  & 10   & 0.005  & 0.01  & < 0.50 \\
                          & 40   & 0.01   & 0.05  & < 0.50 \\
                          & 70   & 0.02   & 0.10  & < 0.50 \\
                          & 100  & 0.03   & 0.15  & 0.63 \\
\ifarXiv\hline\else\midrule\fi
\multirow{4}*{\num{1000}} & 10   & 0.05  & 0.63  & 0.65 \\
                          & 40   & 0.20  & 2.30  & 0.70 \\
                          & 70   & 0.25  & 4.30  & 0.70 \\
                          & 100  & 0.45  & 5.50  & 0.70 \\
\ifarXiv\hline\else\midrule\fi
\multirow{4}*{\num{10000}} & 10   & 0.85   & 13.00   & 5.50 \\
                           & 40   & 3.30   & 120.00   & 6.60 \\
                           & 70   & 5.00   & 250.00   & 7.40 \\
                           & 100  & 6.70   & 335.00   & 8.00 \\
\ifarXiv\hline\else\bottomrule\fi
\end{tabular}
\caption{Maximum CPU and GPU run-time for a single target using the nearest-neighbour estimator and \num{200} surrogates. Summary of the results for different time series lengths ($T=$~\num{100}, \num{1000}, \num{10000}) and different network sizes ($N=$~\num{10}, \num{40}, \num{70}, \num{100}).
}
\label{tb:run-time}
\end{table}

\pagebreak

\subsection{Validation of false positive rate on real fMRI data}
\begin{figure}[htp]
	\centering
	\includegraphics[width=1\textwidth]{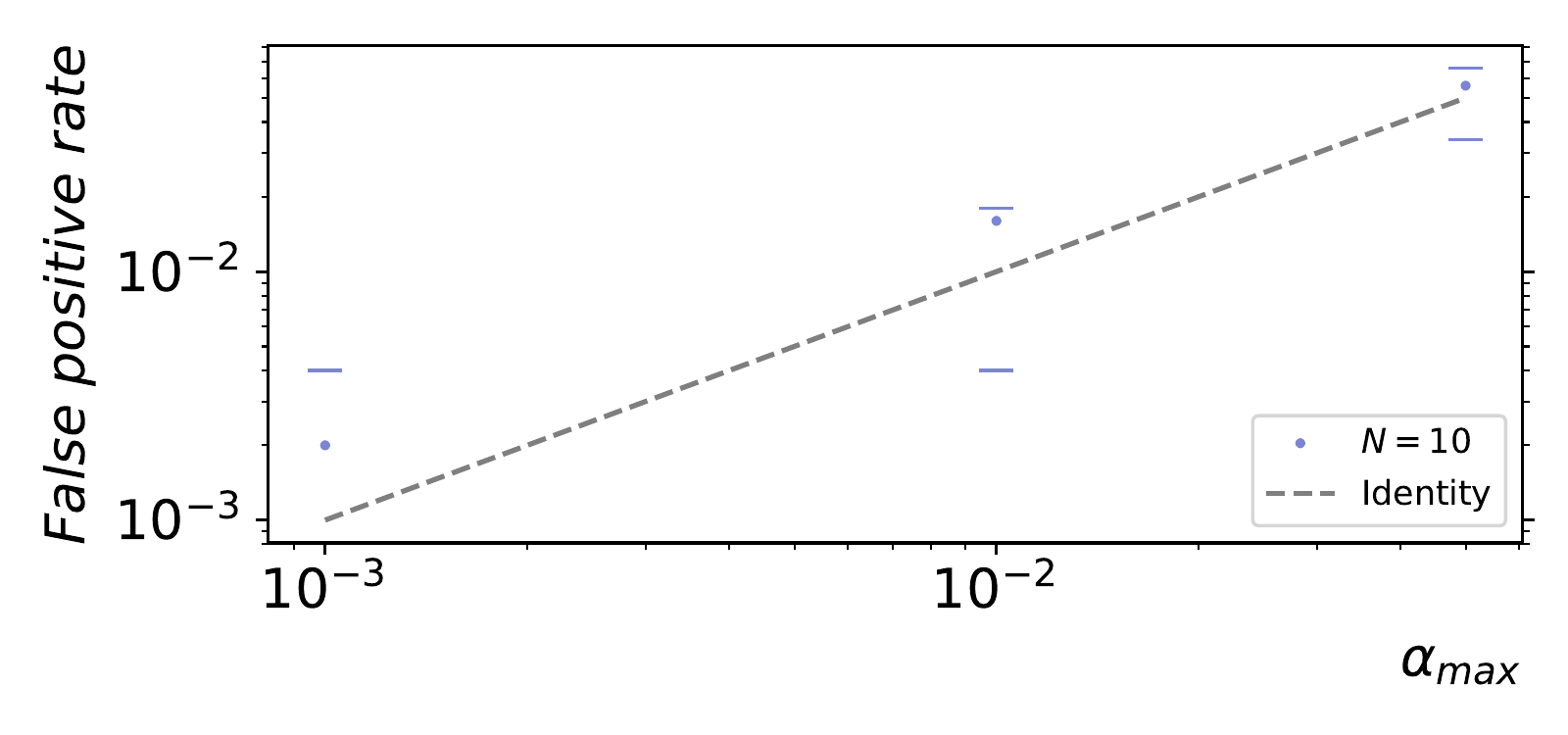}
	 \caption{Validation of false positive rate for a single target ($t_{\textup{FPR}}$) on real fMRI data. The points indicate the average false positive rate over \num{50} repetitions of the experiment using random regions ($N=10$) from different subjects. The horizontal marks indicate the corresponding $5$th and $95$th percentiles of the expected range.
	 These were computed empirically from the distribution of the random variable $\langle X_j/N \rangle$, where $X_j \sim Binomial(N, \alpha_\textup{max})$ are i.i.d. random variables, and the angular brackets indicate the finite average over $50$ repetitions. The identity function is plotted as a reference (dashed line).}
	\label{fig:FPR_HCP}
\end{figure}

The false positive rate validation (presented in \fig{FPR_empty} for synthetic VAR data) was replicated in a scenario where the null hypothesis held for real data. Once again, the aim was to verify that the false positive rate was consistent with the desired level $\alpha_\textup{max}$.
The Human Connectome Project resting state fMRI dataset \citep{VanEssen2012} was used for this purpose (see Supporting Information). The raw data was pre-processed by applying a 3rd order Butterworth bandpass filter ($0.01$-$0.08$ Hz), then cutting $200$ samples from the start and the end of the time series to remove potential filtering artefacts (leaving $800$ samples for the analysis).
In order to build a scenario where the null hypothesis held, $10$ different random regions of interest (ROIs) were selected from different random subjects, such that the corresponding time series were expected to be independent of each other. The network inference was performed with the same settings used in the null test on synthetic data but employing the nearest-neighbour estimator, since the real data could not be assumed to follow a Gaussian distribution. The results on fMRI data are presented in \fig{FPR_HCP} and are consistent with the previous results on synthetic data (\fig{FPR_empty}).

Unless appropriate measures are taken, the strong autocorrelation typically found in real data would result in an inflated false positive rate for short time series (an effect already observed by \citet{Barnett2011} when using Granger causality). The issue is addressed in IDTxl by means of the \emph{dynamic correlation exclusion}, also known as \emph{Theiler window} \citep{Theiler1986, Kantz2003}, as originally suggested for transfer entropy estimation by \citet{Schreiber2000}. The idea is to exclude the closest points in time from the nearest-neighbour search which is necessary for the estimation of the transfer entropy (when using nearest-neighbour estimators). The autocorrelation decay time (\ie,~the shortest time shift such that the autocorrelation function drops by a factor of $1/e$ with respect to the zero-shift value \citep{Lindner2011}) is used as a heuristic to adapt the size of the Theiler window to the data.

\subsection{Alternative visualisations of performance scaling by network and sample size}

\begin{figure}[htp]
	\centering
	\includegraphics[width=0.5\textwidth]{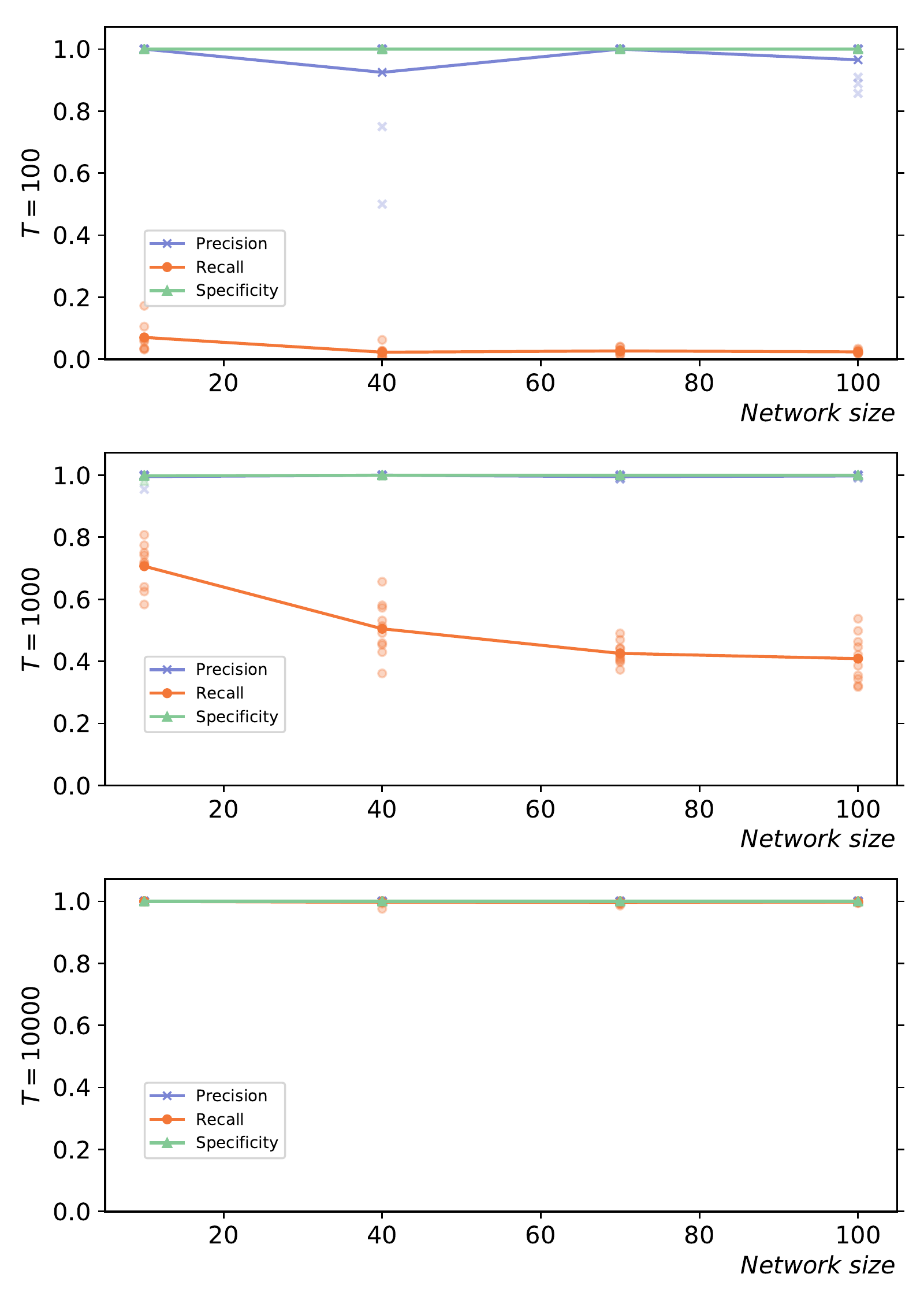}\includegraphics[width=0.5\textwidth]{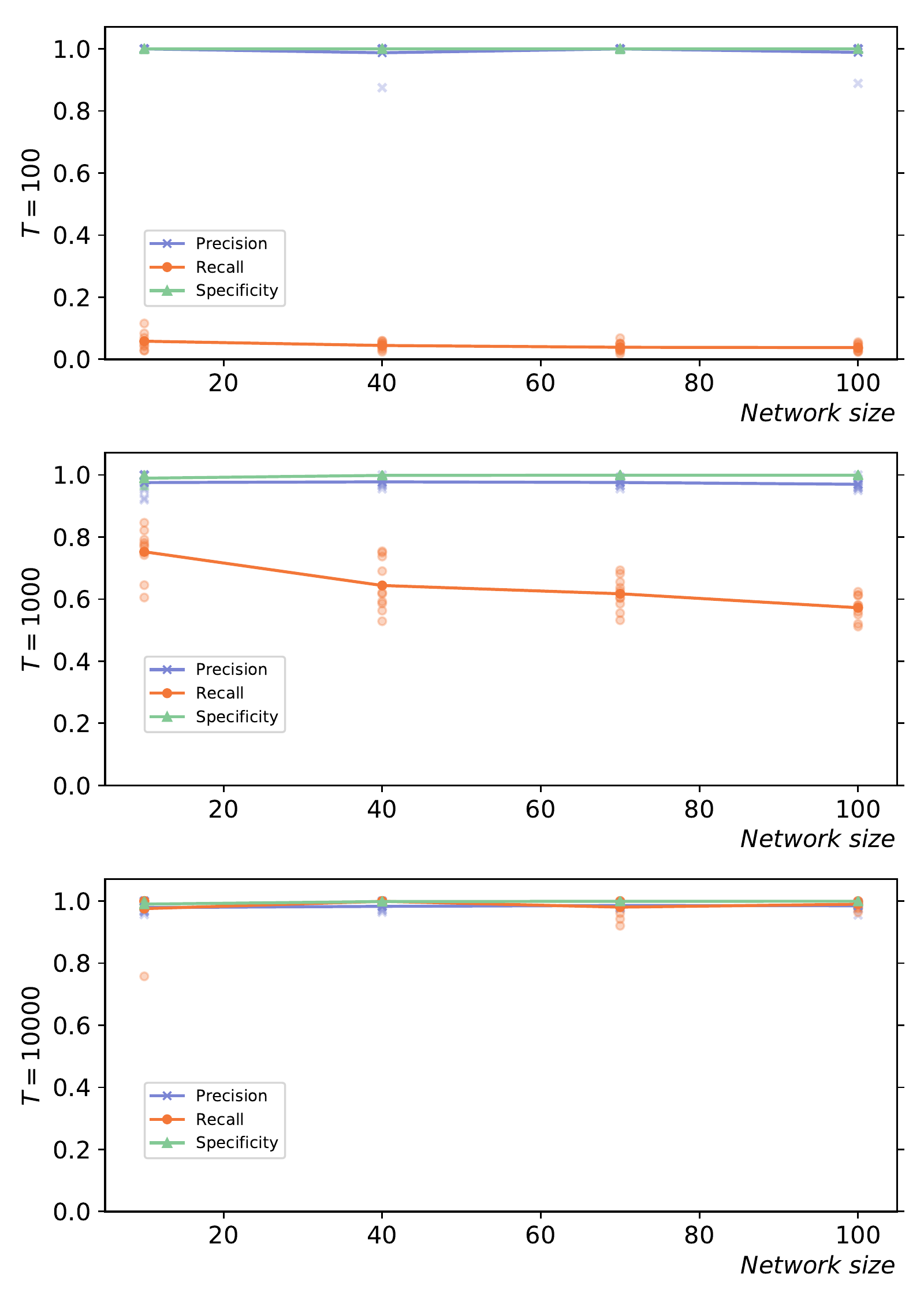}
	\caption{Precision, recall, and specificity for different network sizes, sample sizes, and dynamics. Left: Vector autoregressive process; Right: Coupled logistic maps. Each row corresponds to a different time series length (Top: $T=$~\num{100}; middle: $T=$~\num{1000}, bottom: $T=$~\num{10000}). The results for \num{10} simulations from different initial conditions are shown (low-opacity circles) in addition to the mean values (solid circles).}
	\label{fig:nodes_and_samples_sweep_old}
\end{figure}

\begin{figure}[htp]
	\centering
	\includegraphics[width=0.5\textwidth]{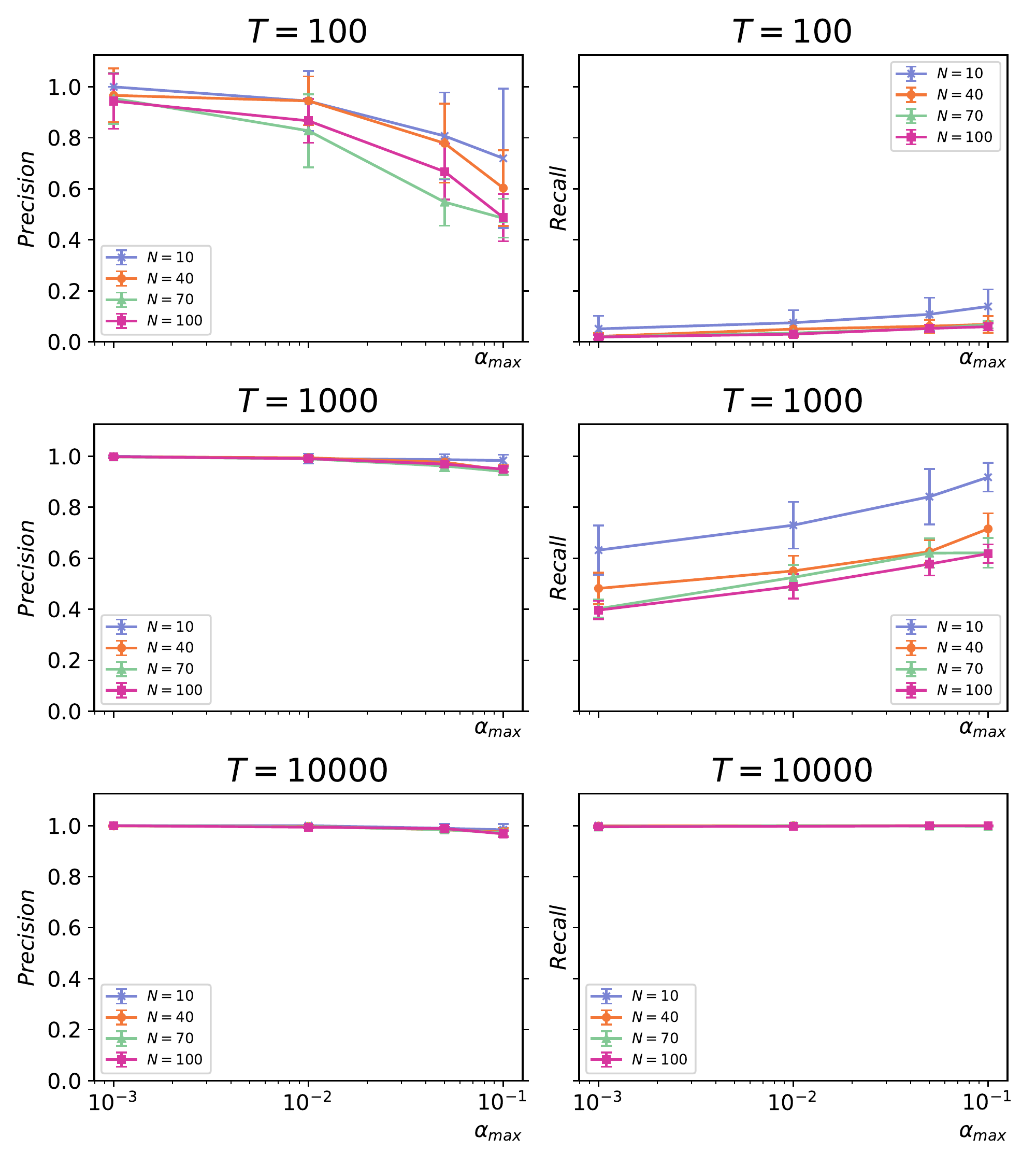}\includegraphics[width=0.5\textwidth]{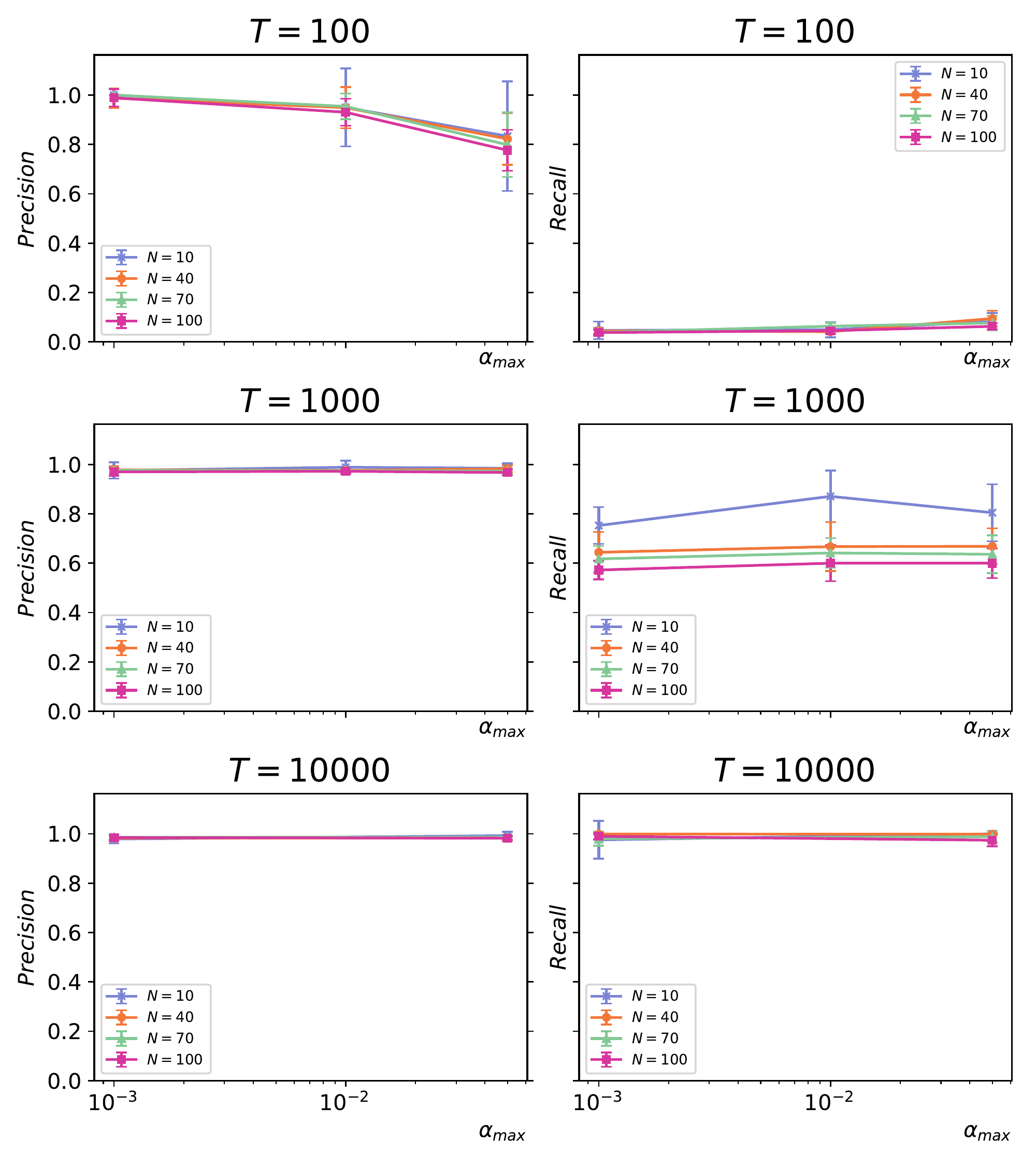}
	\caption{Precision-recall trade-off for different statistical significance levels. The plots show the results for different dynamics (Left: Vector autoregressive process; Right: Coupled logistic maps), different time series lengths (top to bottom: $T=$~\num{100}, \num{1000}, \num{10000}), and different network sizes ($N=$~\num{10}, \num{40}, \num{70}, \num{100}). The error bars indicate the standard deviation over \num{10} simulations from different initial conditions.}
	\label{fig:alpha_sweep}
\end{figure}

\fig{nodes_and_samples_sweep_old} replots the precision and recall from  \fig{nodes_and_samples_sweep} using different subplots for each sample size, and additionally shows the specificity.
Similarly, \fig{alpha_sweep} replots the precision and recall from \fig{alpha_sweep_scatter} using different subplots for each sample size.

\end{document}